\documentclass[aps,showpacs,prb,twocolumn,superscriptaddress]{revtex4}
\usepackage{epsfig}
\usepackage{amsfonts}
\usepackage{amsmath}
\usepackage{mathrsfs, amssymb, amsfonts, amsmath, mathtools}
\usepackage{bm}
\usepackage{float}
\usepackage{subfigure}
\usepackage[utf8]{inputenc}
\usepackage[colorlinks,bookmarks=true,citecolor=blue,linkcolor=blue,urlcolor=blue]{hyperref}
\usepackage[usenames,dvipsnames]{xcolor}

\newcommand{\beq}{\begin{equation}}
\newcommand{\eeq}{\end{equation}}
\newcommand{\bea}{\begin{align}}
\newcommand{\eea}{\end{align}}
\newcommand{\nn}{\nonumber}

\newcommand{\dagga}{{\phantom{\dagger}}}

\begin{document}

\title{Even-odd effects in NSN scattering problems: Application to graphene nanoribbons}
\author{Fran\c{c}ois Cr\'epin}
\affiliation{Institute for Theoretical Physics and Astrophysics,
University of W\"urzburg, 97074 W\"urzburg, Germany}
\author{Hans Hettmansperger}
\affiliation{Institute for Theoretical Physics and Astrophysics,
University of W\"urzburg, 97074 W\"urzburg, Germany}
\author{Patrik Recher}
\affiliation{Institute for Mathematical Physics, TU Braunschweig, 38106 Braunschweig, Germany}
\author{Bj\"orn Trauzettel}
\affiliation{Institute for Theoretical Physics and Astrophysics,
University of W\"urzburg, 97074 W\"urzburg, Germany}

\date{\today}

\begin{abstract}
We study crossed Andreev reflection (CAR) of electrons or holes in normal metal-superconductor-normal metal junctions and highlight some very strong effects of the underlying lattice. In particular, we demonstrate that for sharp interfaces and under certain, albeit generic, symmetry conditions, the CAR probability exactly vanishes for an even number of atoms in the superconducting region. This even-odd effect applies notably to NSN junctions made of graphene nano-ribbons with armchair edges and for zigzag edges with somewhat more restrictive conditions. We analyze its robustness towards smoothing of the boundaries or doping of the sample.
\end{abstract}


\maketitle

\section{Introduction}
A significant amount of work has recently been devoted to the study of possible realizations of spin-entanglers in solid-state devices. Andreev reflection~\cite{Andreev64} at the interface between a normal metal and a superconductor is of particular interest in this context. Indeed the conversion of an incoming hole into a reflected electron at the interface is readily identified with the injection of a Cooper pair into the normal lead and its splitting into two spin-entangled electrons.~\cite{Buttiker03} When two normal leads are connected to a superconducting barrier, a non-local electron-hole conversion \cite{Falci01,Feinberg00} can take place, as a Cooper pair is splitted among both leads. Such a non-local process could be enhanced by Coulomb interaction~\cite{Yeyati07,Lehur06, Recher03, Bena02, Recher02, Recher01, Choi00} or by energy-filtering processes \cite{Cayssol08, Lesovik01}.  Non-local or crossed Andreev reflection (CAR) in normal metal-superconductor-normal metal (NSN) junctions was observed experimentally,~\cite{Mopurgo05,Lohneysen04} and shown to depend strongly on the thickness of the superconductor. CAR is in general not perfect and will coexist with cotunneling, that is, the transfer of an electron or a hole across the superconducting barrier, as well as local normal or Andreev reflection. In non-local conductance measurements, one measures the sum of both effects and it is therefore not possible to measure the CAR signal alone. However, more recently, spatial separation of the CAR signal from the electron cotunneling has been proposed theoretically \cite{Reinthaler12,Yeyati12}. In setups where a bias voltage only exists between the superconductor and the normal metal leads, the electron cotunneling does not appear and one can use current noise measurements to distinguish CAR from the local Andreev reflection \cite{Das12,Wei10} as well as conductance measurements in more involved setups containing quantum dots \cite{Hofstetter10, Hermann10} in the Coulomb blockade regime between the superconductor and the normal metal leads or Aharonov-Bohm rings \cite{Schelter12,Recher01}.

NSN junctions could in principle be made of graphene nanoribbons,~\cite{Wang12, Waintal10, Linder09, Cayssol08} where superconductivity is induced by the proximity to a nearby superconductor.~\cite{Gueron12, Dirks11, Gueron09, Heersche07} At interfaces between graphene and superconductors, new physics emerges because of the peculiar band structure of the carbon monolayer. For instance, in a single NS junction, specular local Andreev reflection was predicted to exist, in striking contrast with conventional metal leads,~\cite{Beenakker06} which has observable effects in the transport characteristics \cite{Beenakker06, Titov07, Cheng11, Schelter12}. In graphene NSN junctions, it appears to be possible to favor CAR over co-tunneling by either considering asymmetrically doped~\cite{Cayssol08} or ferromagnetic leads.~\cite{Linder09, Wang12} { In addition, noise cross correlations in NSN graphene structures have been considered~\cite{Colin08}}. Interestingly, as the length of the superconductor is increased, the CAR probability exhibits oscillations, a consequence of quantum interferences in the superconducting region~\cite{Linder09}, a feature shared by all NSN junctions with interfaces sharp on the scale of the Fermi-wave length in the superconductor.  In our paper, we carefully study these oscillations and go beyond by looking at length dependence on the scale of the lattice constant. In particular, we discover an effect related to the parity of the S region length.\\

\begin{figure}
\centering
\includegraphics[width=7cm,clip]{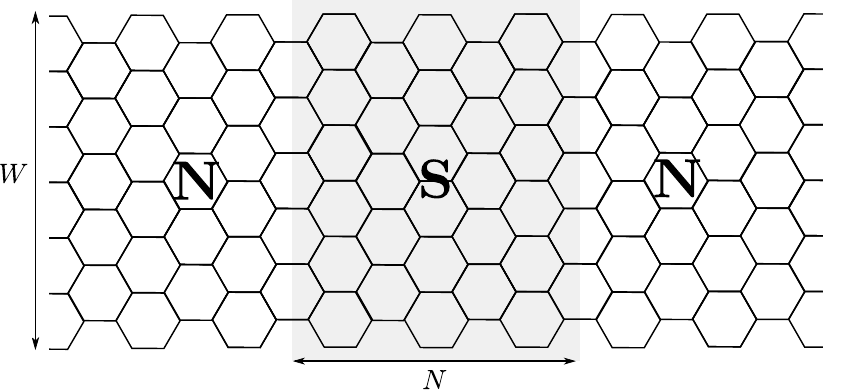}
\caption{ NSN junction made from a graphene nanoribbon with armchair edges. In the S region, an onsite pairing potential $\Delta$ is induced by proximity to an $s$-wave superconductor.}
\label{Fig:NSN_0}
\end{figure}

\begin{figure}
\centering
\includegraphics[width=7.5cm,clip]{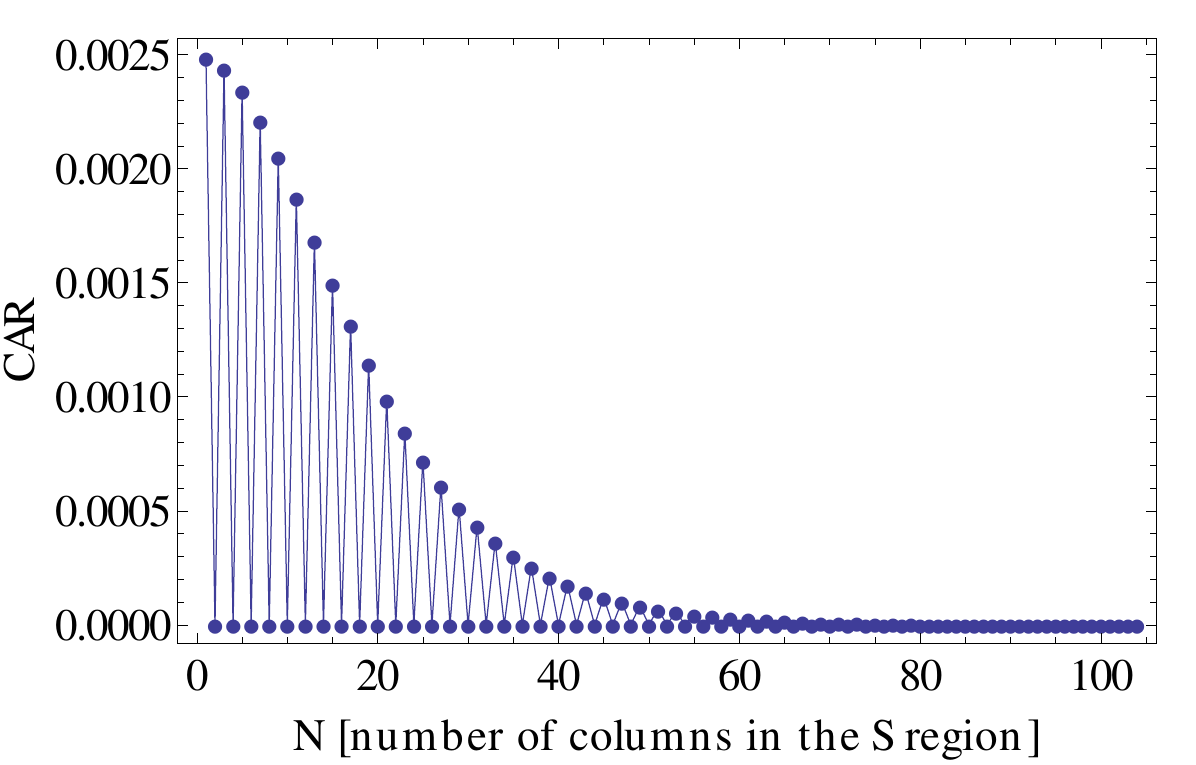}
\caption{(Color online) Crossed Andreev reflection (CAR) probability for the NSN junction of Fig~\ref{Fig:NSN_0}, with $W = 92$, the number of dimers in the transverse direction; the ribbon is metallic. The Fermi level is at the Dirac point in the whole sample. The excitation energy is $\varepsilon = 0.03 t$, and the pairing gap $\Delta = 0.1t$, with $t$ the nearest-neighbor hopping amplitude. Only the metallic mode is excited. The CAR signal vanishes exactly for even $N$.}
\label{Fig:CAR_1}
\end{figure}

This effect is in fact not graphene-specific but appears already in the simplest model of an NSN junction, namely a one-dimensional lattice with nearest-neighbor hopping, where an on-site pairing potential varies from zero in the normal leads to a value $\Delta \neq 0$ in the superconducting region. If the NS interfaces are atomically sharp, as shown in Fig.~\ref{Fig:NSN_1} below, and if the following two conditions are fulfilled
\begin{itemize}
\item[(i)] the spectrum has particle-hole symmetry, that is, the Fermi energy is zero everywhere in the sample and
\item[(ii)] the superconducting region spans an even number of sites,
\end{itemize} 
then the CAR probability vanishes exactly. For an odd number of sites in the S region, the CAR probability does not vanish. To the best of our knowledge, such a strong even-odd effect~\footnote{In the following, {\it even-odd effect} refers to the vanishing of the CAR probability under conditions (i) and (ii).} was not reported before. This effect carries over to armchair graphene nanoribbons, if condition (ii) is interpreted as requiring an even number of transverse zigzag chains in the S region. This situation is represented in Fig.~\ref{Fig:NSN_0} and the CAR probability oscillations, as obtained from wave-function matching, are plotted in Fig.~\ref{Fig:CAR_1}. In armchair nano-ribbons, transverse modes are completely decoupled and their contribution to the CAR signal add up incoherently. It turns out that all transverse modes exhibit the even-odd effect under conditions (i) and (ii). Therefore,  the CAR probability actually vanishes for even lengths of the S region, independent of the excitation energy, below the gap $\Delta$. The zeroes of the CAR probability are analytically understood from an analysis of the symmetries of the problem, at the level of the transfer matrix. In zigzag ribbons the situation is complicated by the strong dependence of Andreev reflection on the ribbon width. \cite{Fazio09} Interestingly, we found the same even-odd effect for anti-zigzag ribbons, yet only for the lowest mode.  

Strikingly, under the symmetry conditions discussed above, the lattice has a strong influence on the transport characteristics, even when the superconducting gap evolves smoothly, but symmetrically, at the two NS-interfaces. Note that such a behavior is not observed in the Bogoliubov-de Gennes equation based on the Dirac equation \cite{Beenakker06} where CAR should vanish when the Fermi energy is zero due to pseudospin conservation. We observe indeed a very small CAR signal in the tight-binding model, however, the even odd effect remains important. The total signal of the CAR could be enhanced significantly in a setup with many modes (note the effect is observed in any armchair mode), this is realized in particular when the ribbon width is large. In the 
case of the zigzag ribbon, the transverse modes will couple in a scattering experiment and the situation is more subtle. 

The article is organized as follows. In Section \ref{sec:linear_chain}, we first demonstrate the even-odd effect for the 1D lattice. In particular, we show that the even-odd effect is independent of the excitation energy, or the strength of the pairing potential. Subsequently, an extension to 2D square lattices is presented. Then, in Section \ref{sec:graphene}, we transpose the analysis to graphene nano-ribbons.  Finally, in Section \ref{sec:extensions}, we discuss non-idealities, in particular doping in the S region and smoothing of the NS interfaces, and interpret numerical results for the CAR probability, obtained with the recursive Green’s function method, in the light of the even-odd effect. Numerical results for the zigzag-ribbon are analysed in the last subsection. Detailed calculations can be found in the appendices.

\section{Even-odd effect in the linear chain}
\label{sec:linear_chain}

\begin{figure}[h] 
\centering
\includegraphics[width=8cm,clip]{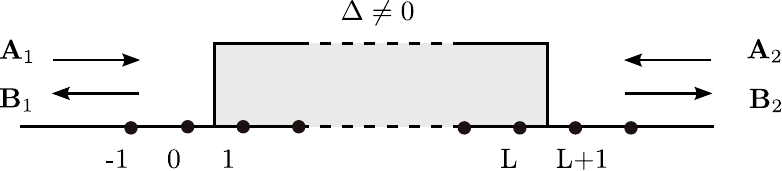}
\caption{NSN junction in a 1D infinite lattice. The S region (shaded box) extends over $L$ sites. }
\label{Fig:NSN_1}
\end{figure} 

We first consider a NSN junction in the simplest one-band model: a chain with nearest-neighbor hopping. The position of a given site is labeled by an integer $\ell$ running from $-\infty$ to $+\infty$ and the S region extends over $L$ sites. We choose, without loss of generality, sites $\ell = 1$ and $\ell = L$ to be the end-sites of the S region, as exemplified in Fig.~\ref{Fig:NSN_1}. Electrons moving on such a lattice, with lattice constant $a$, are then described by the tight-binding Hamiltonian
\beq
H = \sum_{m,n,\sigma} h_{mn}c^\dagger_{m\sigma} c^\dagga_{n\sigma} + \sum_{m,n}\left( \Delta_{mn}c^\dagger_{m\uparrow} c^\dagger_{n\downarrow}+ \textrm{H.c.}\right), \label{eq:H_chain}
\eeq
where $m,n$ run over all integers, $\sigma=\uparrow,\downarrow$ is a spin $1/2$ index and $h_{mn}=-t \delta_{n,m\pm1}$ describes nearest-neighbor hopping while $\Delta_{mn} = \Delta \delta_{m,n}$, for $1\leq m \leq L$, is the pairing potential in the S region. Solving the scattering problem for a given excitation energy $\varepsilon$, that is, finding the scattering matrix $\mathbf{S}(\varepsilon)$, amounts to solving the Bogoliubov-de Gennes (BdG) equations  
\beq
\mathcal{H}_{\textrm{BdG}} \begin{pmatrix}
\Psi_e \\
\Psi_h
\end{pmatrix} = \varepsilon \begin{pmatrix}
\Psi_e \\
\Psi_h
\end{pmatrix}, \label{eq:HBdg}
\eeq
where
\beq
\mathcal{H}_{\textrm{BdG}} =
\begin{pmatrix}
\mathcal{H}_0 -E_F & \mathbf{\Delta} \\
\mathbf{\Delta} & -\mathcal{H}_0 +E_F
\end{pmatrix},
\eeq
with $[\mathcal{H}_0]_{mn} = h_{mn}$ and $[\mathbf{\Delta}]_{mn} = \Delta_{mn}$. $E_F$ is the Fermi energy. We have dropped the spin index due to spin-rotation invariance of the Hamiltonian and chosen $\Delta$ to be real. $\Psi_e$ (resp. $\Psi_h$) describes electron-like (resp. hole-like) excitations. We will look for solutions of the translational invariant problem corresponding to each region (N or S) and match the wave-functions at the NS boundaries.

\subsection{Scattering states and transfer matrix}

\begin{figure}[h]
\centering
\includegraphics[width=7cm,clip]{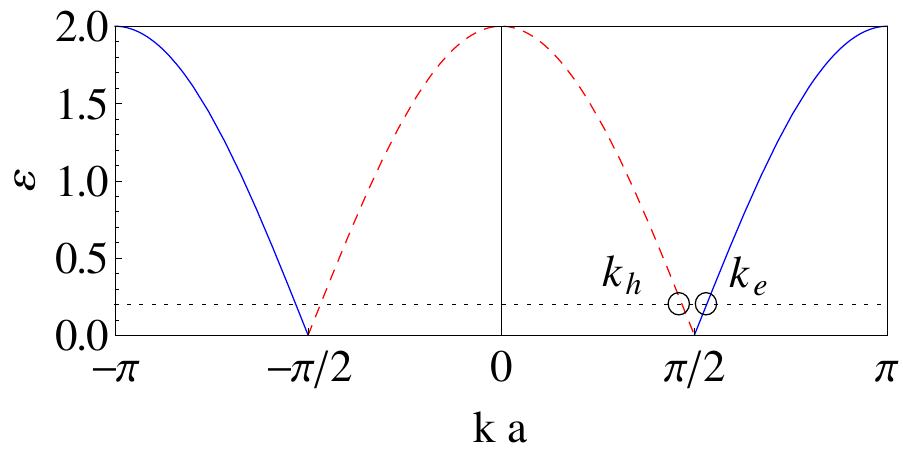}
\caption{(Color online) Spectrum for electron-like (blue, solid) and hole-like (red, dashed) excitations, for $E_F = 0$. We have set $t=a=1$. For a given excitation energy $\varepsilon$ (dotted horizontal line) there are four allowed solutions which correspond to right and left-moving electrons or holes. }
\label{Fig:spectra}
\end{figure}

In the normal regions, where $\Delta_{mn} = 0$, a general solution of Eq.~\eqref{eq:HBdg}, for a given excitation energy $\varepsilon$, is a superposition of incoming and outgoing plane waves,
\begin{align}
\Psi_{e,\ell} &= a_{1,e}  \Psi_{e,\ell}^{+} + b_{1,e}  \Psi_{e,\ell}^{-}, & \ell < 1, \nn \\
\Psi_{h,\ell} &= a_{1,h}  \Psi_{h,\ell}^{+} + b_{1,h}  \Psi_{h,\ell}^{-}, & \ell < 1, \nn \\ 
\Psi_{e,\ell} &= b_{2,e}  \Psi_{e,\ell}^{+} + a_{2,e}  \Psi_{e,\ell}^{-}, & \ell > L, \nn \\
\Psi_{h,\ell} &= b_{2,h}  \Psi_{h,\ell}^{+} + a_{2,h}  \Psi_{h,\ell}^{-}, & \ell > L,\label{eq:ampl} 
\end{align}
where we have defined, for electron-like excitations, right-going waves (+) and left-going waves ($-$) as $\Psi_{e,\ell}^{\pm} = e^{\pm i k_e \ell a}$ while for hole-like excitations, right-going waves and left-going waves are of the form $\Psi_{h,\ell}^{\pm} = e^{\mp i k_h \ell a}$. The wave-vectors $k_e$ and $k_h$ are the positive roots of $\varepsilon = \xi(k)$ and $\varepsilon = -\xi(k)$, respectively, with $\xi(k) = -2t \cos (ka) -E_F$ (see Fig. \ref{Fig:spectra}). Particle-hole symmetry of the spectrum, for $E_F = 0$, implies the important constraint
\beq
k_e + k_h = \pi/a,\label{eq:ke_kh_1}
\eeq
for any excitation energy $\varepsilon$. The amplitudes in Eq.~\eqref{eq:ampl} --- $\mathbf{A}_j = (a_{j,e}, a_{j,h})^{T}$, $\mathbf{B}_j = (b_{j,e}, b_{j,h})^{T}$, for $j=1,2$ --- are related by the scattering matrix $\mathbf{S}$ through
\beq
\begin{pmatrix} \mathbf{B}_1 \\ \mathbf{B}_2 \end{pmatrix} = \mathbf{S} \begin{pmatrix}  \mathbf{A}_1 \\ \mathbf{A}_2  \end{pmatrix}\;. \label{eq:S_1}
\eeq
The scattering matrix is here a $4\times4$ matrix which has the generic form
\beq
\mathbf{S} =
\begin{pmatrix}
\mathbf{r} & \mathbf{t}' \\
\mathbf{t} & \mathbf{r}'
\end{pmatrix}\;, \label{eq:S_2}
\eeq
with $\mathbf{r}, \mathbf{t}, \mathbf{t}', \mathbf{r}'$, $2\times2$ matrices. We find it more convenient to work with transfer matrices -- so-called $M$-matrices -- which relate amplitudes on the right of the scattering potential to amplitudes on the left, that is
\beq
\begin{pmatrix} \mathbf{B}_2 \\ \mathbf{A}_2 \end{pmatrix} = \mathbf{M} \begin{pmatrix}  \mathbf{A}_1 \\ \mathbf{B}_1  \end{pmatrix}\;. \label{eq:M_1}
\eeq
We also introduce a block notation for $\mathbf{M}$ as
\beq
\mathbf{M} =
\begin{pmatrix}
{\bm \alpha} & {\bm \beta} \\
{\bm \gamma} & {\bm \delta}
\end{pmatrix}\;. \label{eq:M}
\eeq
where, again,  ${\bm \alpha}$, ${\bm \beta}$, ${\bm \gamma}$ and ${\bm \delta}$ are $2\times2$ matrices. Some important properties of transfer matrices are summarized in Appendix \ref{sec:appendix2}. In particular, current conservation implies that
\beq \mathbf{t} =  ({\bm \alpha}^\dagger)^{-1}\;. \label{eq:t} \eeq
We show in the next sections that, under the two conditions stated in the introduction, the off-diagonal elements of $ \mathbf{t}$ are zero, and therefore the CAR probability vanishes.


\subsection{Wave-matching at the NS boundaries}

In practice, the transfer matrix as well as the scattering matrix are obtained by searching for plane-wave (possibly with a complex wave-vector) solutions in the superconducting region and matching the wave-functions at the two boundaries. In the S region, plane waves $\Psi_{S,\ell}$ are solutions of
\beq
\begin{pmatrix}
\Lambda(k) & \Delta \\
\Delta & -\Lambda(k)
\end{pmatrix} \Psi_{S,\ell} = \varepsilon \Psi_{S,\ell}
\eeq
with $\Lambda(k) = -2t\cos(ka)-E_F-V_S$, and $V_S$ the doping in the S region. Eigenvalues are $\varepsilon = \pm \sqrt{\Lambda(k)^2 + \Delta^2}$, with eigenvectors $\chi_{\pm}(k) = (F_\pm(k), 1)^T$ and 
\beq
F_\pm(k) = (\Lambda(k)\pm \sqrt{\Lambda(k)^2 + \Delta^2})/\Delta. \label{eq:F}
\eeq 
There are, generically, four solutions for a given excitation energy $\varepsilon$. For $|\varepsilon|>\Delta$, wave-vectors are real, with two right and two left-going waves. However, we are interested in sub-gap transport, that is, $|\varepsilon|<\Delta$, for which electron excitations can be converted into hole excitations, and vice-versa, via the creation or the annihilation of a Cooper pair. In this case, the solutions are four evanescent waves with complex wave-vectors $k$ given by 
\beq
k = \pm \frac{1}{a} \textrm{arccos}\left(-\frac{E_F + V_S \pm i\sqrt{\Delta^2 -\varepsilon^2}}{2 t}\right). \label{eq:k_s1}
\eeq
Two of these wave-vectors have a positive imaginary part and correspond to right-decaying waves while the other two have a negative imaginary part and correspond to left-decaying waves. We define $k_S$ as the wave-vector with positive real and imaginary parts. From Eq.~\eqref{eq:k_s1} it is clear that $-k_S^*$ is the other right-decaying mode while the left-decaying modes are given by $-k_S$ and $k_S^*$. A general solution in the S region is of the form
\begin{align}
\Psi_{S,\ell} &= b_{S,1} \chi_\sigma(k_S) e^{ik_S\ell a} +  b_{S,2} \chi_\sigma(-k_S^ *) e^{-ik_S^*\ell a} \nn \\
&+ a_{S,1} \chi_\sigma(-k_S) e^{-ik_S\ell a} +  a_{S,2} \chi_\sigma(k_S^ *) e^{ik_S^*\ell a}, \label{eq:A,B_S}
\end{align}
with $\sigma = \textrm{sgn}(\varepsilon)$. For a particle-hole symmetric spectrum ($E_F + V_S = 0$) we find the following constraint
\beq 
\mathfrak{Re}(k_S a) = \frac{\pi}{2}.\label{eq:kS_2}
\eeq
The wave-matching conditions~\cite{Schomerus07} read
\beq
\begin{pmatrix}
\Psi_{e,0} \\
\Psi_{h,0}
\end{pmatrix} = \Psi_{S,0} \; \; \textrm{and}\; \;  \begin{pmatrix}
\Psi_{e,1} \\
\Psi_{h,1}
\end{pmatrix} = \Psi_{S,1} \label{eq:wave-matching_1}
\eeq
at the left boundary between sites $0$ and $1$, as well as
\beq
\begin{pmatrix}
\Psi_{e,L} \\
\Psi_{h,L}
\end{pmatrix} = \Psi_{S,L} \; \; \textrm{and}\; \; \begin{pmatrix}
\Psi_{e,L+1} \\
\Psi_{h,L+1}
\end{pmatrix} = \Psi_{S,L+1} \label{eq:wave-matching_2}
\eeq
at the right boundary between sites $L$ and $L+1$. A (numerical) resolution of this system of 8 equations will yield all elements of the transfer or scattering matrices. In the next section, using a suitable decomposition of the transfer matrix, we show analytically, and without solving explicitly the system of equations, how the even-odd effect in the CAR probability is a direct consequence of the symmetries of the problem.

\subsection{Decomposition of the transfer matrix}
\begin{figure}[H] 
\centering
\includegraphics[width=8cm,clip]{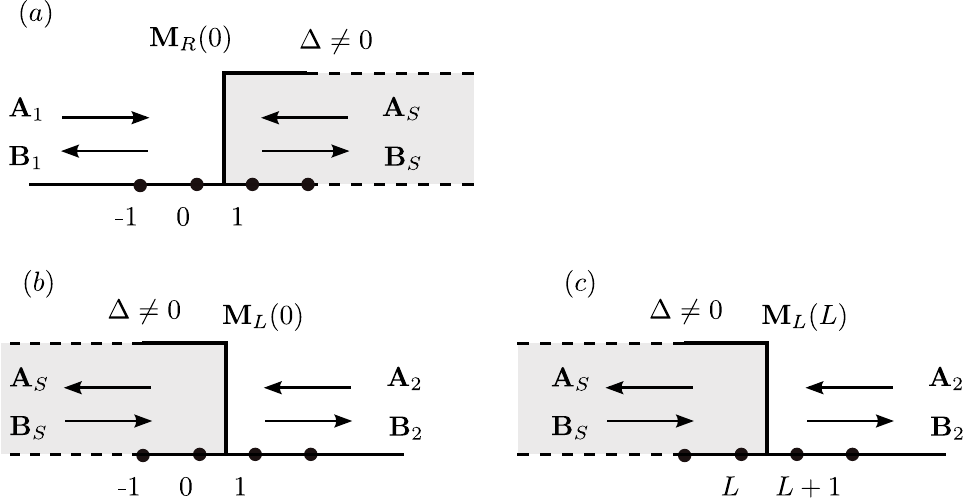}
\caption{Decomposition of the NSN junction of Fig.~\ref{Fig:NSN_1} into 2 successive NS and SN {\it step} junctions represented in diagrams (a) and (c) respectively. Note that in the S region, for $|\varepsilon|<\Delta$, the arrows indicate right or left decaying modes. Diagrams (b) and (c) are related by a translation of $L$ lattice sites. }
\label{Fig:NSN_2}
\end{figure}

One advantage of working with transfer matrices is that upon decomposing the scattering region into several, successive, scattering elements, the total transfer matrix is nothing but the product of transfer matrices for each scattering element. For instance, the NSN junction of Fig.~\ref{Fig:NSN_1} is decomposed into two NS or SN steps, as represented in Fig.~\ref{Fig:NSN_2}. \\

The transfer matrix ${\mathbf{M}}$ of the NSN junction is the product of two transfer matrices, one for each step
\beq
\mathbf{M} = \mathbf{M}_L(L) \mathbf{M}_R(0). \label{eq:M_2}
\eeq
In Fig.~\ref{Fig:NSN_2} we have recasted the amplitudes introduced in Eq.~\eqref{eq:A,B_S} into $\mathbf{A}_S = (a_{S,1}, a_{S,2})^{T}$, $\mathbf{B}_S = (b_{S,1}, b_{S,2})^{T}$. Thus, the transfer matrices $\mathbf{M}_R(0)$ and $\mathbf{M}_L(L)$ relate amplitudes in the S region to amplitudes on the left and on the right, respectively, through the equations
\beq
\begin{pmatrix} \mathbf{B}_S \\ \mathbf{A}_S \end{pmatrix} = \mathbf{M}_R(0) \begin{pmatrix}  \mathbf{A}_1 \\ \mathbf{B}_1  \end{pmatrix}, \label{eq:M_3}
\eeq
and
\beq
\begin{pmatrix} \mathbf{B}_2 \\ \mathbf{A}_2 \end{pmatrix} = \mathbf{M}_L(L) \begin{pmatrix}  \mathbf{B}_S \\ \mathbf{A}_S  \end{pmatrix}. \label{eq:M_4}
\eeq
Equation \eqref{eq:M_2} is a direct consequence of Eqs. \eqref{eq:M_1}, \eqref{eq:M_3} and \eqref{eq:M_4}. Note that Eqs.~\eqref{eq:M_3}, \eqref{eq:M_4} are only reformulations of the wave-matching conditions Eqs. \eqref{eq:wave-matching_1}, \eqref{eq:wave-matching_2}, respectively. Interestingly, there is a simple relation between $\mathbf{M}_R(0)$ and $\mathbf{M}_L(L)$. First note that the junctions (c) and (b) in Fig.~\ref{Fig:NSN_2} are related by a translation of $L$ lattice sites. Therefore, amplitudes for the two associated scattering problems only differ by phase factors, that is
\begin{align}
\mathbf{M}_L(L) = \textrm{Diag}[e^{-i k_e L a}, e^{i k_h L a}, e^{i k_e L a}, e^{-i k_h L a}]\nn \\
\times \mathbf{M}_L(0)\textrm{Diag}[e^{i k_S L a}, e^{-i k_S^* L a}, e^{-i k_S L a}, e^{i k_S^* L a}]. 
\end{align}
Next we work out the relation between $\mathbf{M}_L(0)$ and $\mathbf{M}_R(0)$. Due to the left-right symmetry of the problem, we also have
\beq
\mathbf{M}_L(0) = [\mathbf{M}_R(0)]^{-1}. \label{eq:M_5}
\eeq
As we did for ${\mathbf{M}}$, we introduce a block decomposition for  $\mathbf{M}_L(0)$ and $\mathbf{M}_R(0)$ through
\beq
\mathbf{M}_{R(L)}(0) =
\begin{pmatrix}
{\bm \alpha}_{R(L)} & {\bm \beta}_{R(L)} \\
{\bm \gamma}_{R(L)} & {\bm \delta}_{R(L)} 
\end{pmatrix}. \vspace{0.2cm} \label{eq:M_6}
\eeq
Following Eq.~\eqref{eq:t}, we compute the upper-left block ${\bm \alpha}$ in $\mathbf{M}$. We factor out the free propagation and introduce
\beq
{\bm \alpha} = \begin{pmatrix}
e^{-i k_e L a} & 0 \\
0 & e^{i k_h L a}
\end{pmatrix}  \tilde{{\bm \alpha}}.
\eeq
From Eqs.~\eqref{eq:M_2}, \eqref{eq:M_5}, \eqref{eq:M_6} and using blockwise inversion formulae, ${\bm \alpha}_L = \left[{\bm \alpha}_R -{\bm \beta}_R {\bm \delta}_R^{-1} {\bm \gamma}_R\right]^{-1} $ and  ${\bm \beta}_L =- \left[{\bm \alpha}_R -{\bm \beta}_R {\bm \delta}_R^{-1} {\bm \gamma}_R\right]^{-1}{\bm \beta}_R {\bm \delta}_R^{-1} $, we arrive at
\begin{widetext}
\beq
\tilde{{\bm \alpha}}^{-1} = e^{i k_S L a} \left[ e^{i 2 k_S L a} 
\begin{pmatrix}
1 & 0 \\
0 & e^{-i 2 \mathfrak{Re}(k_S) L a}
\end{pmatrix}
- {\bm \alpha}_R^{-1} {\bm \beta}_R {\bm \delta}_R^{-1} \begin{pmatrix}
1 & 0 \\
0 & e^{i 2 \mathfrak{Re}(k_S) L a}
\end{pmatrix} {\bm \gamma}_R  \right]^{-1} \left[ \mathbf{1}_2 - {\bm \alpha}_R^{-1}{\bm \beta}_R {\bm \delta}_R^{-1} {\bm \gamma}_R\right]. \label{eq:alpha_tilde_1}
\eeq
\end{widetext}
So far, we have only used the left-right symmetry of the barrier, and Eq.~\eqref{eq:alpha_tilde_1} is very general. To prove our statement about the CAR probability we now simplify the expression for $\tilde{{\bm \alpha}}^{-1}$ using the conditions (i) $E_F = V_S = 0$ and (ii) $L$ even, stated earlier. Recalling that particle-hole symmetry of the spectrum implies $\mathfrak{Re}(k_S a) = \pi/2$ --- see Eq.~\eqref{eq:kS_2} --- it is now appropriate to use the decomposition 
\beq\displaystyle \begin{pmatrix}
1 & 0 \\
0 & e^{i\pi L}
\end{pmatrix} = \frac{1+e^{i\pi L}}{2} \mathbf{1}_2 + \frac{1-e^{i\pi L}}{2} \sigma_z, \label{eq:sigma_z}
\eeq
from which it follows that if $L$ is an even integer then
\beq
\tilde{{\bm \alpha}}^{-1}= e^{i k_S L a} \left[ (e^{i 2 k_S L a}-1)\mathcal{A}^{-1} +
 \mathbf{1}_2 
 \right]^{-1}, \label{eq:alpha_even}
\eeq 
where we have introduced the matrix
\beq
\mathcal{A} =  \mathbf{1}_2 - \left({\bm \alpha}_R^{-1}{\bm \beta}_R\right)\left( {\bm \delta}_R^{-1} {\bm \gamma}_R\right) = \mathbf{1}_2 -  {\bm r}_L' {\bm r}_R. \label{eq:mathcalA}
\eeq
In the last equation,  ${\bm r}_R$  and ${\bm r}_L'$ are two unitary matrices, that can be identified with the reflection matrices of the NS steps of Fig~\ref{Fig:NSN_2}(a) and Fig~\ref{Fig:NSN_2}(b), respectively, see App. \ref{sec:appendix2}. To prove that the CAR amplitude is zero, it is enough to prove that the matrix $\mathcal{A}$ has zero off-diagonal elements.  

\begin{figure}[h!] 
\centering
\includegraphics[width=8cm,clip]{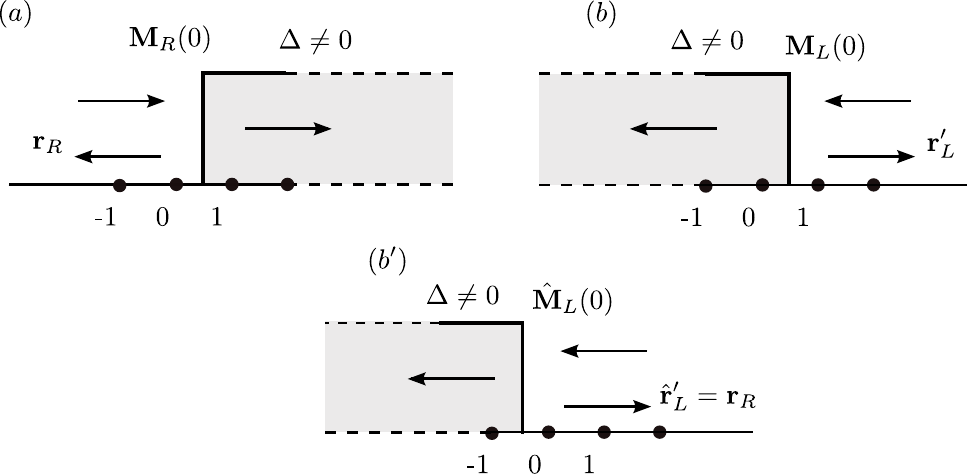}
\caption{The NS junction in (b') is the same as the one in (b), only shifted by one lattice site to the left. Is is related to the junction in (a) by a parity transformation. }
\label{Fig:NSN_3b}
\end{figure}

The two matrices ${\bm r}_R$  and ${\bm r}_L'$ are related by a translation of one lattice site. This is illustrated on Fig.~\ref{Fig:NSN_3b}, where we represented the NS junction (b') which is related to (b) by a translation of one site and to (a) by a parity transformation -- see App. \ref{sec:appendix2} for a detailed proof. As a consequence,
\beq
\mathcal{A} = \mathbf{1}_2 - \begin{pmatrix}
e^{-ik_e a} & 0 \\
0 & e^{ik_h a}
\end{pmatrix} \mathbf{r}_R \begin{pmatrix}
e^{-ik_e a} & 0 \\
0 & e^{ik_h a}
\end{pmatrix} \mathbf{r}_R. \label{eq:mathcalA_2}
\eeq
Particle-hole symmetry of the spectrum also implies $\mathbf{r}_R = \sigma_x \mathbf{r}_R \sigma_x$. It follows that off-diagonal elements of $\mathcal{A}$ are zero. Indeed we have:
\beq
\mathcal{A}^{eh} = e^{-2ik_e a}\left(1 + e^{i(k_h+k_e)a}\right)r_R^{eh} r_R^{ee} = 0,
\eeq
and
\beq
\mathcal{A}^{he} =e^{2ik_h a}\left(1 + e^{-i(k_h+k_e)a}\right)r_R^{he} r_R^{hh} = 0,
\eeq
due to the constraint in Eq.~\eqref{eq:ke_kh_1}.

\subsection{Extension to the 2D square lattice}
The extension of the present study to the 2D square lattice is straightforward. We consider an NSN junction on a ribbon of width $W$. $a$ is again the lattice constant and $t$ is the hopping energy in both longitudinal and transverse directions. Imposing open boundary conditions, the transverse wave-vector $p$ is quantized as
\beq
p = \frac{r \pi}{(W + 1)a}, \; \; \; \; \textrm{with} \; \; r = 1,2,3, \ldots
\eeq
and is independent of the longitudinal wave-vector $k$. The 2D scattering problem decomposes into several independent 1D scattering problems, one for each available transverse mode at a given excitation energy $\varepsilon$. The spectrum of the ribbon is
\beq
\xi(k,p) = -E_F -2t\cos(p a) -2t\cos(k a).
\eeq
This means that each transverse mode has an effective Fermi energy $E_F' = E_F +2t\cos(pa)$, that can in principle be tuned to zero. For such modes, with $E_F' =  0$, one will observe the same even-odd effect as in the linear chain, namely the absence of CAR for even lengths of the superconducting region. But for an arbitrary excitation energy, many propagating modes with $E_F' \neq 0$ are likely to be excited -- this number grows with the width of the ribbon -- { and as their contributions add up incoherently in the CAR signal, fast oscillations due to the even-odd effect will be masked}.

In the next section, we will see that armchair graphene nanoribbons, due to their peculiar (sub-)lattice geometry, exhibit an even-odd effect at zero Fermi energy $E_F$, for each and every transverse mode.

\section{Even-odd effect in graphene nanoribbons}
\label{sec:graphene}

{
We now turn to the case of NSN junctions in graphene nano-ribbons. We make the assumption of proximity induced superconductivity and study the behavior of the CAR probability close to the charge neutrality point $E_F = 0$. Note that, contrary to 2D graphene, the density of states at $E_F = 0$ is non zero in both metallic armchair and zig-zag ribbons.~\cite{Wakabayashi10}}

\subsection{Armchair edges}

\subsubsection{Bogoliubov-de Gennes equations}
\begin{figure} 
\centering
\includegraphics[width=7cm,clip]{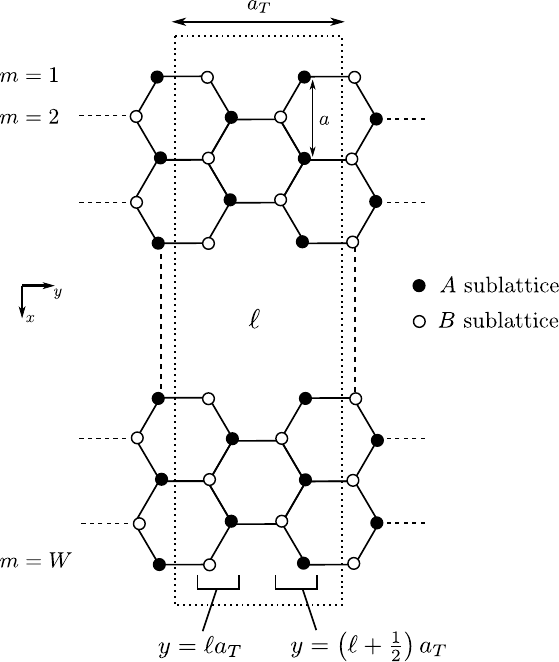}
\caption{Unit-cell of an armchair graphene nanoribbon of width $W$. An atom has coordinates $(m a/2, y=y_{\ell,mA(B)})$, where $\ell$ is the index of the unit-cell and $A, B$ is the sublattice index. }
\label{Fig:graphene_unit_cell}
\end{figure}

We now consider a graphene nanoribbon with armchair edges, as represented in Figs.~\ref{Fig:graphene_unit_cell} and \ref{Fig:NSN_4}. Let $t$ be the hopping energy between neighboring lattice sites, $a$ the distance between sites on a given sublattice and $a_T = \sqrt{3} a$ the length of a unit-cell. The width of the ribbon is $W$ and counts the number of dimer lines in the transverse direction. We follow closely the notation of Ref.~\onlinecite{Wakabayashi10}. Let us write down the tight binding Hamiltonian $H = H_{0\uparrow}+H_{0\downarrow} + H_{\textrm{pair}}$ for armchair graphene ribbons, with
\begin{align}
H_{0\sigma} = &-t\sum_{\ell}\left[\sum_{m\;\textrm{even}} a^\dagger_{\ell,m,\sigma}b^\dagga_{\ell,m,\sigma} + \sum_{m\;\textrm{odd}} a^\dagger_{\ell,m,\sigma}b^\dagga_{\ell+1,m,\sigma}\right] \nn \\
&-t\sum_{\ell}\sum_{m} a^\dagger_{\ell,m,\sigma}b^\dagga_{\ell,m+1,\sigma} + b^\dagger_{\ell,m,\sigma}a^\dagga_{\ell+1,m+1,\sigma}  \nn \\ &  +\textrm{H.c.}
\end{align}
for $\sigma = \uparrow,\downarrow$ and
\beq
H_{\textrm{pair}} = \sum_{\ell,m} \Delta_{\ell m} \left[ a^\dagger_{\ell,m,\uparrow} a^\dagger_{\ell,m,\downarrow} + b^\dagger_{\ell,m,\uparrow} b^\dagger_{\ell,m,\downarrow}\right] + \textrm{H.c.} \;.
\eeq
Here, $a^\dagger_{\ell,m,\sigma}$ and $b^\dagger_{\ell,m,\sigma}$ are creation operators on the $A$ and $B$ sublattices, respectively, while $a^\dagga_{\ell,m,\sigma}$ and $b^\dagga_{\ell,m,\sigma}$ are the conjugate annihilation operators. We look for plane-wave solutions in the normal region, by taking the Fourier transform of the Hamiltonian in the $y$ direction. Furthermore,  spin-rotation invariance of the problem allows us to drop the spin 1/2 index. Plane-wave solutions can thus be written as
\beq
\Psi_{\ell,m}(k) =  \left(\varphi_{A}(k)e^{i k y_{\ell,mA}} + \varphi_{B}(k)e^{i k y_{\ell,mB}} \right) \sin\left(  m \frac{a}{2} p\right), \label{eq:wave_graphene}
\eeq
where $\left( \varphi_{A}(k), \varphi_{B}(k) \right)^T$ is a spinor of the form
\beq
\begin{pmatrix}
\varphi_{A}(k)\\
\varphi_{B}(k)
\end{pmatrix} =  \begin{pmatrix}
-s e^{-i\theta(k)} \\
1
\end{pmatrix},  \label{eq:spinor_1}
\eeq
and the phase $\theta(k)$ is defined in Eq.~\eqref{eq:phase} below. Several comments are in order.  Due to open boundary conditions on the armchair edges, the transverse momentum $p$ is quantized as
\beq
p = \frac{2 r \pi}{(W + 1)a} \; \; \; \; \textrm{with} \; \; r = 1,2,3, \ldots
\eeq
We furthermore use the convention that 
\begin{align}
&y_{\ell,1A} = y_{\ell,2B} = y_{\ell,3A} = \ldots \equiv \ell a_T, \\
&y_{\ell,1B} = y_{\ell,2A} = y_{\ell,3B} = \ldots \equiv (\ell+1/2) a_T,
\end{align}
as indicated in Fig.~\ref{Fig:graphene_unit_cell}. In other words we absorb the differences of coordinates between sites $mA$ and $(m+1)B$ in the phase $\theta(k)$. Spinors in Eq.~\eqref{eq:spinor_1} are eigenstates of the effective 2-band Hamiltonian

\beq
\mathcal{H}_0(k,p) = \begin{pmatrix}
0 & \epsilon_p + e^{-i k a_T/2} \\
\epsilon_p + e^{i k a_T/2} & 0
\end{pmatrix}\label{eq:H_2band}
\eeq
with eigenvalues $E_s(k,p) = s|\epsilon_p + e^{-ika_T/2}|$. We have defined $\epsilon_p=2\cos(p a/2)$ and $s=\pm 1$ is a band index. One can check that the phase difference between the spinor components is given by
\beq
e^{-i\theta(k)} =  \left(\frac{\epsilon_p + e^{-i k a_T/2}}{\epsilon_p + e^{i k a_T/2}}\right)^{1/2}. \label{eq:phase}
\eeq 

Note that, including the pairing potential, the Bogolubov-de Gennes equations reduce to the following effective form 
\beq
\begin{pmatrix}
\mathcal{H}_0(k,p)-E_F & \Delta \mathbf{1}_2\\
\Delta \mathbf{1}_2 & -\mathcal{H}_0(k,p)+E_F
\end{pmatrix}\begin{pmatrix}
\Psi_e \\
\Psi_h
\end{pmatrix} = \varepsilon \begin{pmatrix}
\Psi_e \\
\Psi_h
\end{pmatrix} ,
\eeq
and can be solved independently for each value of the transverse momentum $p$, at a given excitation energy $\varepsilon$. $\Psi_e$ and $\Psi_h$ are now spinors, and $\mathcal{H}_0(k)$ is the effective 2-band Hamiltonian introduced in Eq.~\eqref{eq:H_2band}.

\subsubsection{Spectrum and scattering states}
\begin{figure} 
\centering
\includegraphics[width=8cm,clip]{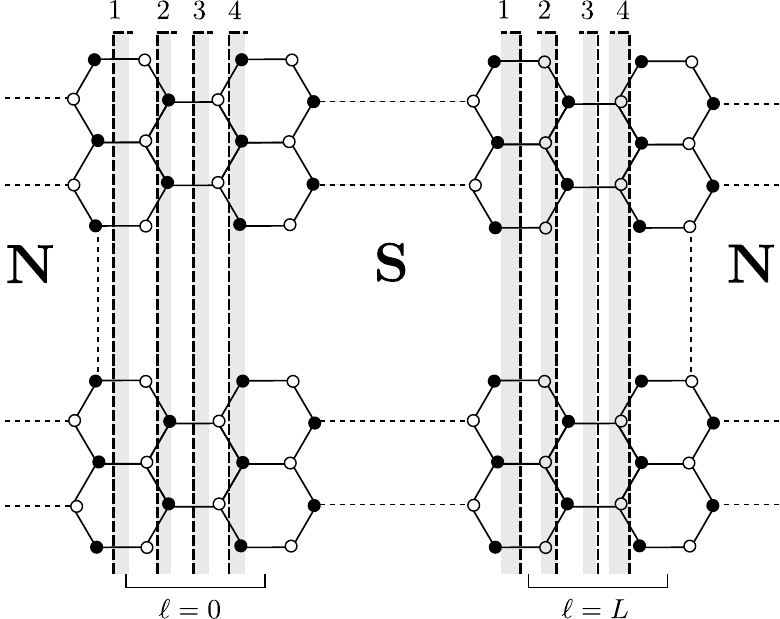}
\caption{NSN junction on an armchair graphene nanoribbon. The S region spans $L$ unit-cells, in the sense that the left boundary is in the $0^{\textrm{th}}$ unit-cell while the right boundary is in the $L^{\textrm{th}}$ unit-cell. In each of these two cells there are two inequivalent positions for the NS junction, 1 and 2 or 3 and 4 (see main text).}
\label{Fig:NSN_4}
\end{figure}

The superconducting region where $\Delta$ is non-zero extends over $L$ unit-cells, in the sense that the left NS interface lies in the cell $\ell =0$ while the right interface  lies in the cell $\ell = L$, as pictured in Fig.~\ref{Fig:NSN_4}. Anticipating what follows and building on the formalism introduced for the treatment of the linear chain, we write the transfer matrix $\mathbf{M}^{(i,j)}$ of such an NSN junction as a product of two simpler transfer matrices, $\mathbf{M}_R^{(i)}(0)$ and $ \mathbf{M}_L^{(j)}(L)$, one for each NS interface. Thus,
\beq
\mathbf{M}^{(i,j)}= \mathbf{M}_L^{(j)}(L) \mathbf{M}_R^{(i)}(0).
\eeq
The labels $i$ and $j$ can, in principle, take the values $1,2,3,4$ corresponding to four distinct positions of the interfaces, as depicted in Fig.~\ref{Fig:NSN_4}. However only positions 1 and 2 are inequivalent, as 3 and 1 are related by a translation of half the unit-cell; the same is true for 2 and 4. Due to the two sublattices in graphene, 1 and 2 are not related by a simple translation in the longitudinal direction. In this section, we restrict $i$ and $j$ to the values 1 and 2. There are in total, for fixed $L$, four a priori inequivalent configurations of the NSN junction, which are labeled by $(i,j) = (1,1), (2,2), (1,2)$, or $(2,1)$. The condition of an even number of sites in the S region here translates to $(i,j) = (1,1)$ or $(2,2)$, which actually corresponds to an even number of zigzag columns. We will show that under conditions (i) particle-hole symmetry of the spectrum, that is, $E_F=V_S=0$ and (ii) an even number of sites in the S region, the CAR probability vanishes, for each transverse mode. Incidently, if condition (i) is fulfilled then the odd configurations $(1,2)$ and $(2,1)$ are equivalent, a consequence of time-reversal symmetry. For these two configurations, the CAR probability is finite.\\

\begin{figure}[h!]
\centering
\includegraphics[width=8cm,clip]{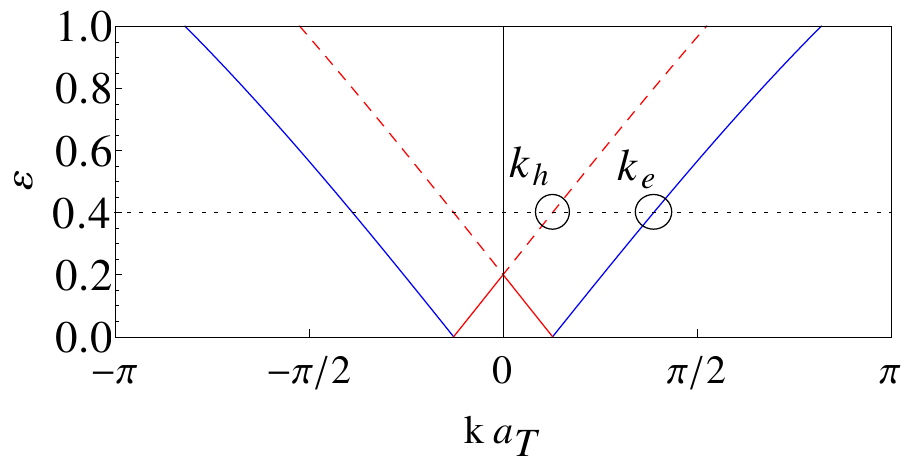}
\caption{(Color online) Electron-like (blue) and hole-like (red) excitations for the metallic mode of armchair graphene ($W=32$), with $E_F = 0.2$. For $\varepsilon = 0.4$ (horizontal dotted line), electron excitation are in the conduction band (solid line, $s=1$) and holes in the valence band (dashed line, $s'=-1$).}
\label{Fig:spectra3}
\end{figure}

First, let us define the scattering states and state the wave-matching conditions. We work for a given transverse mode, with momentum $p$. Similar to the 1D lattice, we introduce scattering states as superpositions of incoming and outgoing plane-waves according to 
\begin{align}
\Psi_{e,\ell,m} &= a_{1,e}  \Psi_{e,\ell,m}^{+} + b_{1,e}  \Psi_{e,\ell,m}^{-},  \nn \\
\Psi_{h,\ell,m} &= a_{1,h}  \Psi_{h,\ell,m}^{+} + b_{1,h}  \Psi_{h,\ell,m}^{-}, \label{eq:A,B_2a}
\end{align}
in the left region, and
\begin{align}
\Psi_{e,\ell,m} &= b_{2,e}  \Psi_{e,\ell,m}^{+} + a_{2,e}  \Psi_{e,\ell,m}^{-},  \nn \\
\Psi_{h,\ell,m} &= b_{2,h}  \Psi_{h,\ell,m}^{+} + a_{2,h}  \Psi_{h,\ell,m}^{-}, \label{eq:A,B_2b}
\end{align}
in the right region. The right ($+$) and left ($-$) going wave-functions are defined in App. \ref{sec:appendix1}. Their wave-vectors are given by the solutions of $\varepsilon = -E_F + E_s(k,p)$ and $\varepsilon = E_F - E_{s'}(k,p)$, for electron-like and hole-like excitations, respectively. We define $k_e$ and $k_h$ as the positive roots of these equations. An example is given in Fig.~\ref{Fig:spectra3}. Particle-hole symmetry of the spectrum, $E_F = 0$, implies 
\beq
k_e = k_h. \label{eq:ke_kh_2}
\eeq


In the S region, we take into account possible doping that shifts the Fermi energy by $V_S$. The BdG Hamiltonian has there four eigenvalues and eigenvectors. For a given excitation energy, there are again four possible solutions. As far as sub-gap transport is concerned, we write the corresponding wave-vectors as $k_S$, $-k_S^*$ (right-decaying modes) and $-k_S$, $k_S^*$ (left-decaying modes) where $k_S$ is the solution with positive real and imaginary parts (see App. \ref{sec:appendix1} for more details). In the particle-hole symmetric case, $E_F + V_S = 0$, the wave-vectors satisfy the important constraint
\beq
\mathfrak{Re}(k_S) = \left\lbrace \begin{array}{ll} 0 &\mbox{if $\epsilon_p<0$}, \\  2\pi/a_T &\mbox{if $\epsilon_p>0$}.   \end{array} \right.
\eeq

A general solution of the BdG equations in the S region is  a superposition of right and left decaying modes of the form
\begin{align}
\Psi_{S,\ell,m} &= b_{S,1} \mathfrak{D}(k_S)\chi_{1+}(k_S)  +  b_{S,2} \mathfrak{D}(-k_S^*) \chi_{1+}(-k_S^ *)  \nn \\
&+ a_{S,1} \mathfrak{D}(-k_S)\chi_{1+}(-k_S) +  a_{S,2} \mathfrak{D}(k_S^*)\chi_{1+}(k_S^ *), \label{eq:A,B_S_2}
\end{align}
where we have introduced the diagonal matrix $\mathfrak{D}(k) = \textrm{Diag}[e^{iky_{\ell,mA}},e^{iky_{\ell,mB}},e^{iky_{\ell,mA}},e^{iky_{\ell,mB}}]$, and  $\chi_{1+}$ is a wave-vector of $\mathcal{H}_{\textrm{BdG}}$. Evidently, Eqs.~\eqref{eq:A,B_2a}, \eqref{eq:A,B_2b} and \eqref{eq:A,B_S_2} are, up to complications due to the sublattice, analogous to Eqs. \eqref{eq:ampl} and \eqref{eq:A,B_S} for the 1D lattice. The sublattice degree of freedom really enters into the problem in the possibility of having inequivalent interfaces. This is made more formal by writing down the wave-matching conditions. For a type 1 boundary, they read

\beq
\begin{pmatrix}
\Psi_{e,\ell-1,m}^{A} \\
\Psi_{e,\ell,m}^{B} \\
\Psi_{h,\ell-1,m}^{A} \\
\Psi_{h,\ell,m}^{B} 
\end{pmatrix} = 
\begin{pmatrix}
\Psi_{S,\ell-1,m}^{1} \\
\Psi_{S,\ell,m}^{2} \\
\Psi_{S,\ell-1,m}^{3} \\
\Psi_{S,\ell,m}^{4} 
\end{pmatrix}
\eeq
with $\ell = 0,L$ and for all $m$. For a type 2 boundary,
\beq
\begin{pmatrix}
\Psi_{e,\ell,m}^{A} \\
\Psi_{e,\ell,m-1}^{B} \\
\Psi_{h,\ell,m}^{A} \\
\Psi_{h,\ell,m-1}^{B} 
\end{pmatrix} = 
\begin{pmatrix}
\Psi_{S,\ell,m}^{1} \\
\Psi_{S,\ell,m-1}^{2} \\
\Psi_{S,\ell,m}^{3} \\
\Psi_{S,\ell,m-1}^{4} 
\end{pmatrix}
\eeq
with $\ell = 0,L$ and for all $m$. These four sets of equations define the transfer matrices, $\mathbf{M}_R^{(1)}(0)$, $ \mathbf{M}_L^{(1)}(L)$, $\mathbf{M}_R^{(2)}(0)$ and $ \mathbf{M}_L^{(2)}(L)$.

\subsubsection{Decomposition of the transfer matrix and even odd effect}

We now express $\mathbf{M}^{(i,j)}$ in terms of matrices in the unit-cell $0$ only. Using the transformation rule for a translation of $L$ unit-cells of length $a_T$, we write
\begin{align}
\mathbf{M}^{(i,j)} &=  \textrm{Diag}[e^{-ik_e L a_T},e^{-ik_h L a_T},e^{ik_e L a_T},e^{ik_h L a_T}] \nn \\
&\times \mathbf{M}_L^{(j)}(0)  \nn \\
&\times \textrm{Diag}[
e^{ik_S L a_T} ,e^{-ik_S^* L a_T} ,e^{-ik_S L a_T} , e^{ik_S^* L a_T}] \nn \\ &\times\mathbf{M}_R^{(i)}(0). 
\end{align}
We introduce the following block notation for these matrices,
\beq
\mathbf{M}^{(i,j)} =
\begin{pmatrix}
{\bm \alpha} & {\bm \beta} \\
{\bm \gamma} & {\bm \delta}
\end{pmatrix}, 
\mathbf{M}_{R(L)}^{(i)}(0) = \begin{pmatrix}
{\bm \alpha}_{R(L)}^{(i)} & {\bm \beta}_{R(L)}^{(i)} \\
{\bm \gamma}_{R(L)}^{(i)} & {\bm \delta}_{R(L)}^{(i)} \end{pmatrix}.
\eeq
To analyze the CAR processes, we have to compute $\mathbf{t} =  ({\bm \alpha}^\dagger)^{-1}$. We first factor out the free propagation by introducing
\beq
{\bm \alpha} = \begin{pmatrix}
e^{-i k_e L a_T} & 0 \\
0 & e^{-i k_h L a_T}
\end{pmatrix}  \tilde{{\bm \alpha}}
\eeq 
and using blockwise inversion formulae, we arrive at
\begin{widetext}
\begin{align}
 \tilde{{\bm \alpha}}^{-1} = & 
 e^{i k_S L a_T} \left[ e^{i 2 k_S L a_T} 
\begin{pmatrix}
1 & 0 \\
0 & e^{-i 2 \mathfrak{Re}(k_S) L a_T}
\end{pmatrix}
- [{\bm \alpha}_R^{(i)}]^{-1} {\bm \beta}_R^{(j)} [{\bm \delta}_R^{(j)}]^{-1} \begin{pmatrix}
1 & 0 \\
0 & e^{i 2 \mathfrak{Re}(k_S) L a_T}
\end{pmatrix} {\bm \gamma}_R^{(i)}  \right]^{-1}\nn \\
& \hspace{6cm}\times [{\bm \alpha}_R^{(i)}]^{-1} {\bm \alpha}_R^{(j)} \left[ \mathbf{1}_2 - [{\bm \alpha}_R^{(j)}]^{-1}{\bm \beta}_R^{(j)} [{\bm \delta}_R^{(j)}]^{-1} {\bm \gamma}_R^{(j)}\right]. \label{eq:transmission_graphene}
\end{align}
\end{widetext}

\begin{figure}[h!] 
\centering
\includegraphics[width=8cm,clip]{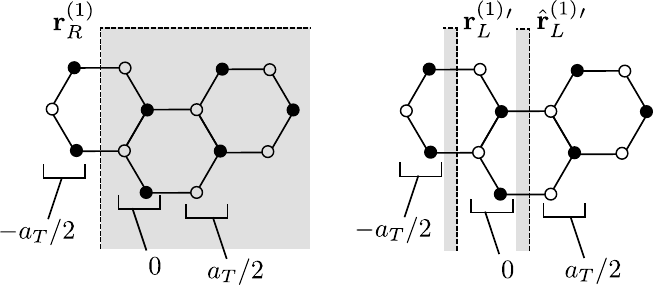}
\caption{$\mathbf{r}_L^{(1)}{}'$ and $\hat{\mathbf{r}}_L^{(1)}{}'$ are related by a translation of half a unit-cell. Due to the sublattice degree of freedom, $\hat{\mathbf{r}}_L^{(1)}{}'$ and $ {\bm r}_R^{(i)}$ are not just related by a parity transformation but instead through phase factors (see Eq.~\eqref{eq:r_graphene}).}
\label{Fig:NSN_5b}
\end{figure}

At this stage we ask for conditions (i) and (ii) to be true. As we said earlier, condition (ii), an even number of sites, is equivalent to the left-right symmetry $i=j$. Condition (i) implies in the case of graphene $\mathfrak{Re}(k_S a_T) = 0$ or $2\pi$. Hence, for even lengths of the S region,
\begin{align}
 \tilde{{\bm \alpha}}^{-1} = e^{i k_S L a_T} \left[ (e^{i 2 k_S L a_T}-1)\mathcal{A}^{-1} +
 \mathbf{1}_2 
 \right]^{-1} 
\end{align}
with
\beq
\mathcal{A} = \mathbf{1}_2 - [{\bm \alpha}_R^{(i)}]^{-1} {\bm \beta}_R^{(i)} [{\bm \delta}_R^{(i)}]^{-1} {\bm \gamma}_R^{(i)}= \mathbf{1}_2 -  {\bm r}_L^{(i)}{}' {\bm r}_R^{(i)},
\eeq
where ${\bm r}_L^{(i)}{}'$ and ${\bm r}_R^{(i)}$ are two unitary reflection matrices, on the right and left interfaces respectively. Similarly to the 1D case, we are left to prove that $\mathcal{A}$ has zero off-diagonal elements. To do so we try to find a simple relation between ${\bm r}_L^{(i)}{}'$ and $ {\bm r}_R^{(i)} $.  We take $i = 1$ for now, but a similar proof holds for $i=2$. ${\bm r}_L^{(1)}{}'$ and $ {\bm r}_R^{(1)} $ are related by a translation of half a unit-cell {\it plus} a sublattice exchange as is apparent in Fig.~\ref{Fig:NSN_5b}. We find
\begin{align}
\mathbf{r}_L^{(1)}{}' = \textrm{Diag}[
-se^{isk_e a_T/2}e^{-is\theta_e}, -s'e^{-is'k_h a_T/2}e^{is'\theta_h}]
\mathbf{r}_R^{(1)} \nn \\
\times \textrm{Diag}[
-se^{isk_e a_T/2}e^{-is\theta_e}, -s'e^{-is'k_h a_T/2}e^{is'\theta_h}], \label{eq:r_graphene}
\end{align}
for which a detailed proof is given in Appendix \ref{sec:appendix2}. Due to particle-hole symmetry of the spectrum, the unitary matrix $\mathbf{r}_R^{(1)}$ satisfies $\mathbf{r}_R^{(1)} = \sigma_x \mathbf{r}_R^{(1)} \sigma_x$. The off-diagonal elements of $\mathcal{A}$ are given by
\begin{align}
\mathcal{A}^{eh} &= r^{(1)eh}_R r^{(1)ee}_R \left(  e^{isk_e a_T}e^{-2is\theta_e} \right. \nn \\ & \left. +  s s'e^{i(sk_e a_T/2-s'k_h a_T/2)} e^{-i(s\theta_e - s'\theta_h)} \right) \end{align}
and
\begin{align}
\mathcal{A}^{he} &= r^{(1)he}_R r^{(1)hh}_R\left(  e^{-is'k_h a_T}e^{-2is\theta_h} \right. \nn \\ & \left. + s s'e^{i(sk_e a_T/2-s'k_h a_T/2)} e^{-i(s\theta_e - s'\theta_h)} \right)\;.
\end{align}
Particle-hole symmetry implies $k_e = k_h$ -- see Eq.~\eqref{eq:ke_kh_2} -- and in turn $\theta_e = \theta_h$, as well as $ss' = -1$. This leads to zero off-diagonal elements of $\mathcal{A}$, and then of $\mathbf{t}$. Hence, the CAR probability vanishes. 

\subsection{zigzag edges}

{ We now briefly turn to the case of graphene ribbons with zigzag edges. Whereas one can treat armchair ribbons as quasi-1D wires, the inter-dependence of longitudinal and transverse momenta in zigzag ribbons makes such an approach difficult. We therefore computed the scattering matrix numerically, using a recursive Green's function algorithm.~\cite{Lee81,MacKinnon85,Todorov94,Lake97,Schelter10} We focus on the lowest mode, in order to simplify the analysis by avoiding inter-mode scattering. The main difference with respect to armchair graphene is a strong dependence on the parity of the width of the ribbon.~\cite{Fazio09, Wang12} For so-called zigzag edges (an even number of chains in the transverse direction) and zero Fermi energy, local (resp. crossed) Andreev reflection is suppressed in the lowest mode, due to opposite parities of the transverse wave-functions of incoming electrons and reflected (resp. transmitted) holes.~\cite{Fazio09, Wang12} For anti-zigzag edges~\cite{Fazio09, Akhmerov08} however, local Andreev reflection is allowed at zero Fermi energy, as incoming electrons and reflected holes have now the same parity. We found that for sharp NS interfaces, although incoming electrons and transmitted holes have still opposite parities, CAR is not necessarily suppressed since the pairing potential breaks the sublattice symmetry. In that case, an even-odd effect is indeed present, as shown in Fig.~\ref{Fig:zz1}.

\begin{figure}[h!]
\centering
\includegraphics[width=8cm,clip]{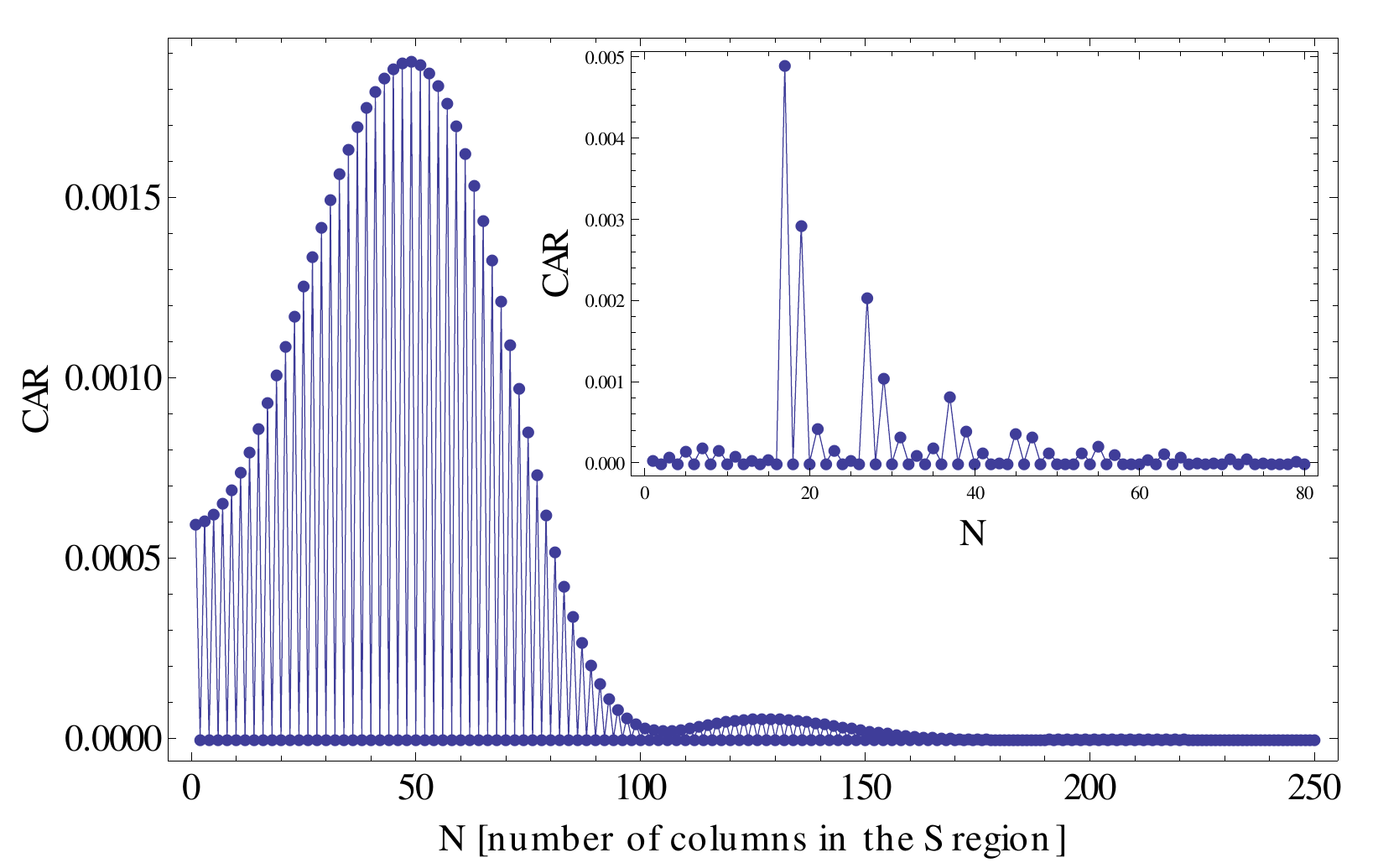}
\caption{(Color online) CAR probability in an anti-zigzag ribbon, for $E_F = V_S = 0$, $\Delta = 0.0031 t$ and $\varepsilon = 0.003 t$, and $W=3$ (number of chains). The inset shows the CAR probability for a wider ribbon with $W=21$, all other parameters kept equal.  }
\label{Fig:zz1}
\end{figure}

Qualitatively, electrons and holes in the lowest mode just propagate at the edges of the ribbon and therefore the situation is similar to the 1D chain. The oscillations in the signal corresponding to the odd configurations is nevertheless unique to graphene with zigzag edges.

}

\section{Non-idealities}
\label{sec:extensions}

\subsection{Armchair edges}

The even-odd effect reported in the present paper relies on two stringent conditions, particle-hole symmetry of the spectrum, that is, $E_F = V_S = 0$, and atomically sharp NS interfaces. There are several ways to relax these two conditions and we inspect them in turn, in the case of an armchair graphene nanoribbon. We use, for that purpose, two numerical methods. Either we compute elements of the scattering matrix by matching waves at the interface, numerically -- this will be done for the simpler cases -- or using a recursive Green's functions algorithm.~\cite{Lee81,MacKinnon85,Todorov94,Lake97,Schelter10} In all the examples we consider a metallic ribbon of width $W = 92$, and probe lengths of the S region up to a hundred $a_0$, with $a_0$ the distance between carbon atoms, which amounts to a few nanometers. Energies are in units of $t \simeq 3 eV$, the nearest neighbor hopping. The superconducting gap is chosen to be $\Delta = 0.1 t$. For excitation energies below $E_{g} = 0.058 t$, only one mode, the metallic mode, is probed and contributes to the signal. This is the case in all plots in this section, where $\varepsilon = 0.03 t$. These are rather high values, as compared to experiments. Indeed the induced gap is usually a few m$eV$ ($\sim 0.001 t$).~~\cite{Dirks11, Gueron09, Heersche07} However the superconducting gates have typical lengths $L_S$ of a few hundred nanometers.~\cite{Dirks11, Heersche07} Therefore, with values realized in experiments, the dimensionless ratio $\Delta L_S/\hbar v_F $ is of similar order as the same ratio computed with the values used in our calculation.

Note that, for a given length $N$ (measured in number of columns) of the S region, there are several possible configurations of the NS interfaces. We choose the following sweep: we fix the boundary in the unit-cell 0 to be in configuration $i=1$ and sweep the four configurations $j=1, 2, 3, 4$ in the unit-cell $L$.\\

\subsubsection{Effects of doping}

\begin{figure}[h!]
\centering
\includegraphics[width=8cm,clip]{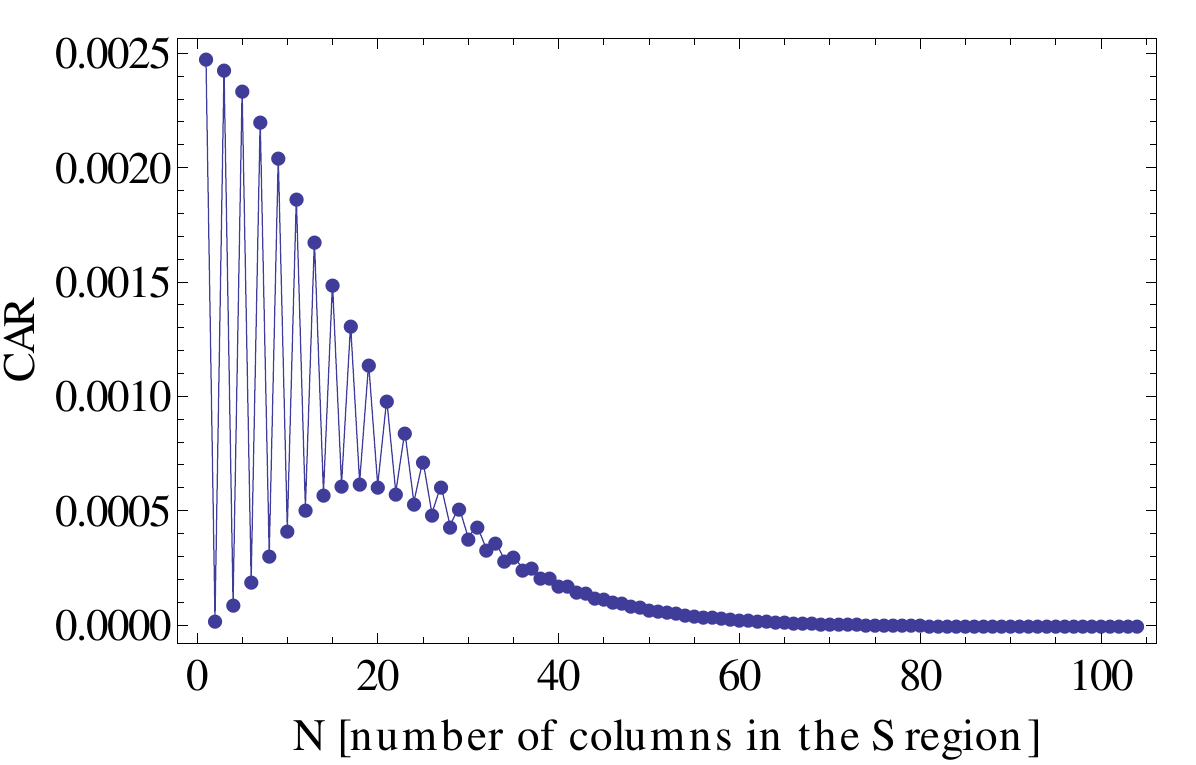}
\caption{(Color online) CAR probability for $E_F = 0.1 t$, $V_S = -E_F$, $\Delta = 0.1 t$ and $\varepsilon = 0.03t$, obtained by wave-function matching. Even and odd configurations are still clearly distinct. }
\label{Fig:extension0}
\end{figure}

In this section we keep interfaces sharp. Let us look, for pedagogical purpose, at the following situation, $E_F \neq 0$ and $V_S = -E_F$. This has two consequences. First, the real part of $k_S$ in the S region remains zero, hence we still expect a strong difference between even and odd configurations, as can be inferred from Eq. \eqref{eq:transmission_graphene}.  However, particle-hole symmetry is broken and the off-diagonal elements of the matrix $\mathcal{A}$ are not zero anymore. Then, the CAR probability does not vanish for even lengths of the S region, as can be seen in Fig.~\ref{Fig:extension0}.

\begin{figure}[h!]
\centering
\includegraphics[width=8cm,clip]{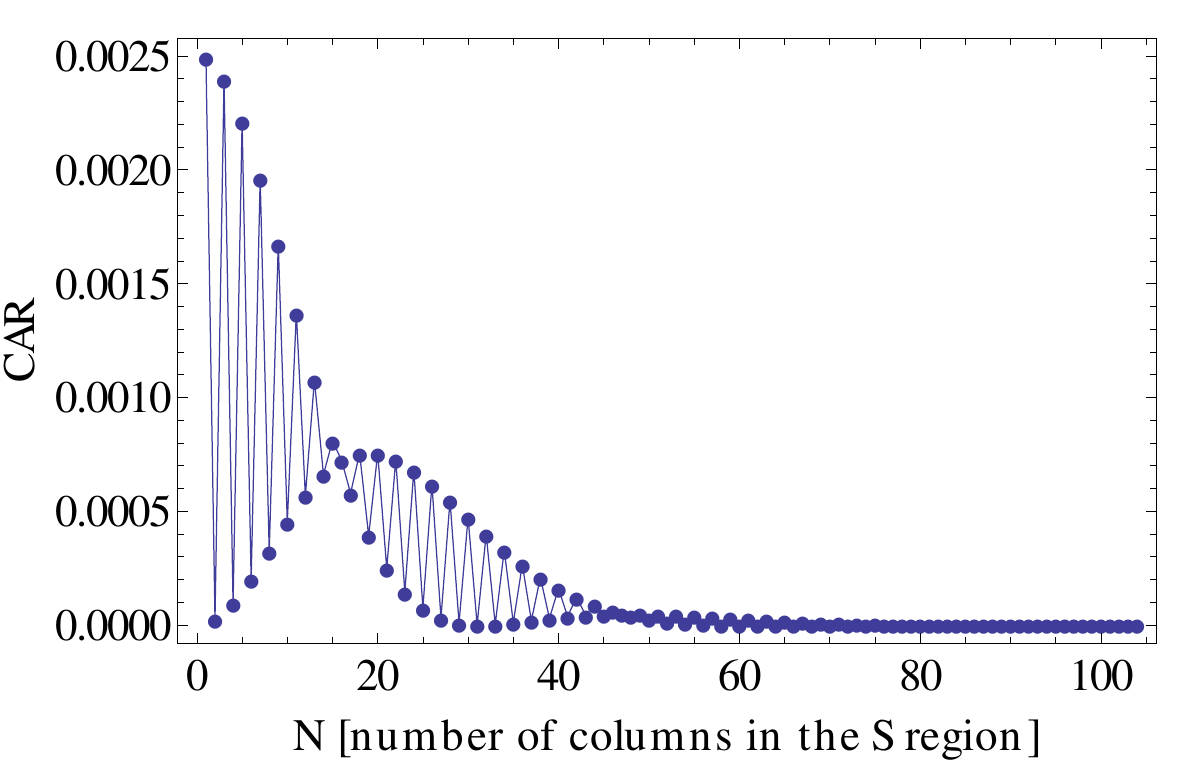}
\caption{(Color online) CAR probability for $E_F = 0.1 t$, $V_S = 0$, $\Delta = 0.1 t$ and $\varepsilon = 0.03t$, obtained by wave-function matching. }
\label{Fig:extension1}
\end{figure}

If we now take $E_F\neq 0, V_S = 0$, that is, a constant doping throughout the sample, then the wave-vector $k_S$ acquires a non-zero real part, and the two curves start oscillating, as can be seen in Fig.~\ref{Fig:extension1}.

\begin{figure}[h!]
\centering
\subfigure{
\includegraphics[width=8cm,clip]{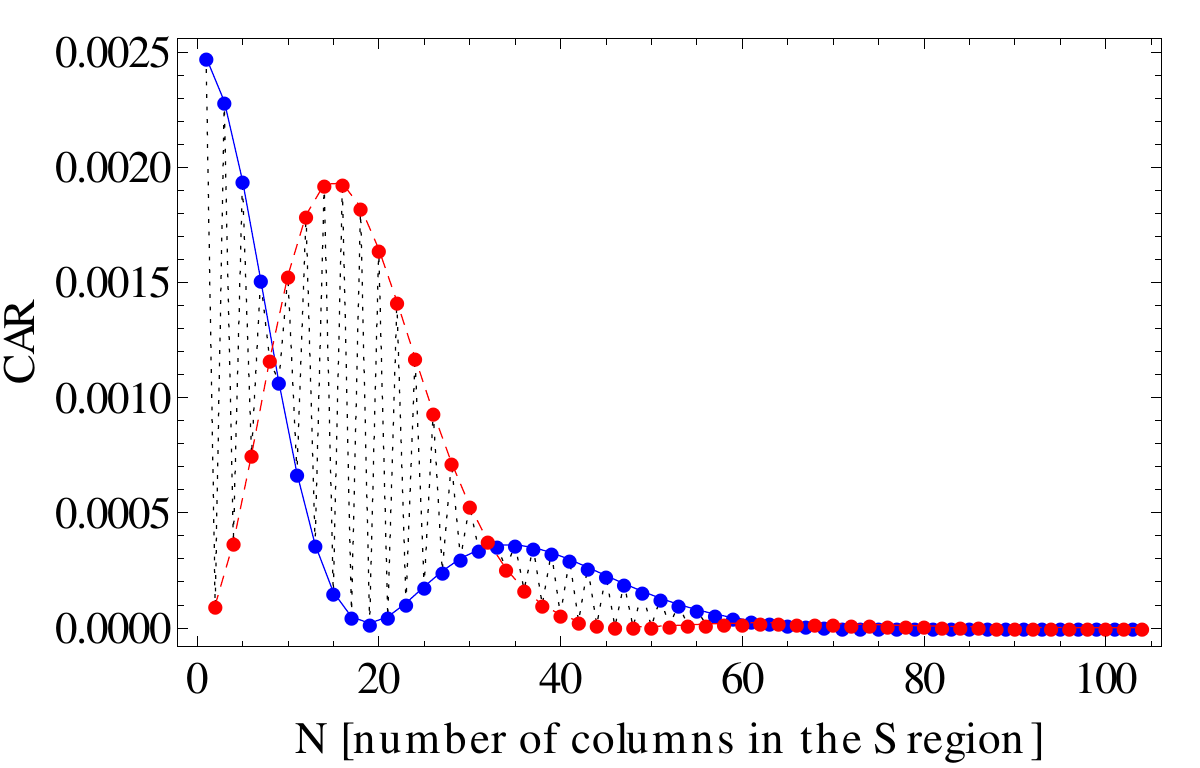}}
\vspace{0.5cm}
\subfigure{
\includegraphics[width=8cm,clip]{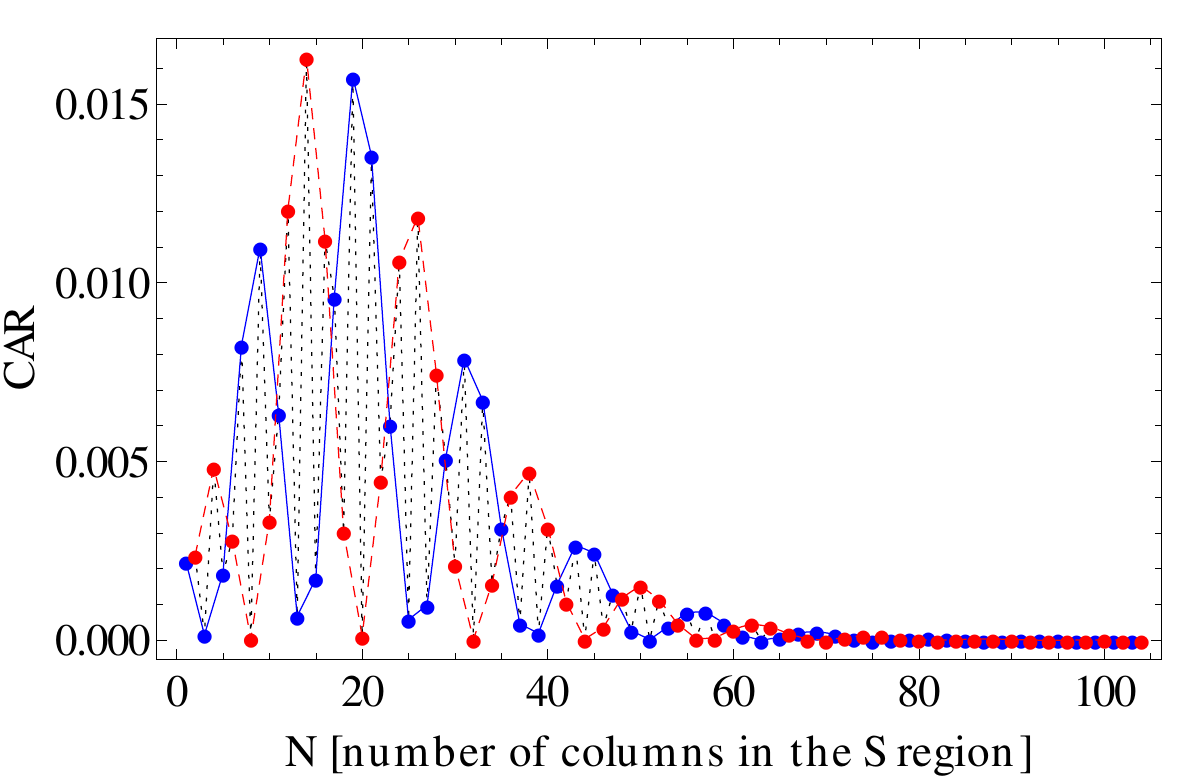}}
\caption{(Color online) CAR probability for $E_F = 0$, $V_S = 0.1 t$, $\Delta = 0.1 t$, $\varepsilon = 0.03t$ (top panel) and $E_F = 0$, $V_S = 0.5 t$, $\Delta = 0.1 t$, $\varepsilon = 0.03t$ (bottom panel). The red dashed line (resp. blue solid line) corresponds to even (resp. odd) configurations, and shows a long-wave length modulation due to doping. The signal oscillates fastly between the two curves (dotted line), a reminiscence of the even-odd effect.  }
\label{Fig:extension2}
\end{figure}

Next, we look at the case where there is only doping in the superconducting region, $E_F = 0, V_S \neq 0$. As can be seen in Fig.~\ref{Fig:extension2}, the signal will, in general, show oscillations similar to the previous case of doping the whole sample. { These long wave-length Fabry-Perot oscillations are ubiquitous in scattering problems, where the scattering region has a finite extent. In the context of superconductivity they are reminiscent of the well-known Tomasch effect in continuous models of NSN junctions, where oscillations with wavenumber $k_F$ arise due to  multiple  reflections  of gapped quasiparticles  at the sample-boundaries.~\cite{Burstein1969} }

\begin{figure}[h!]
\centering
\includegraphics[width=8cm,clip]{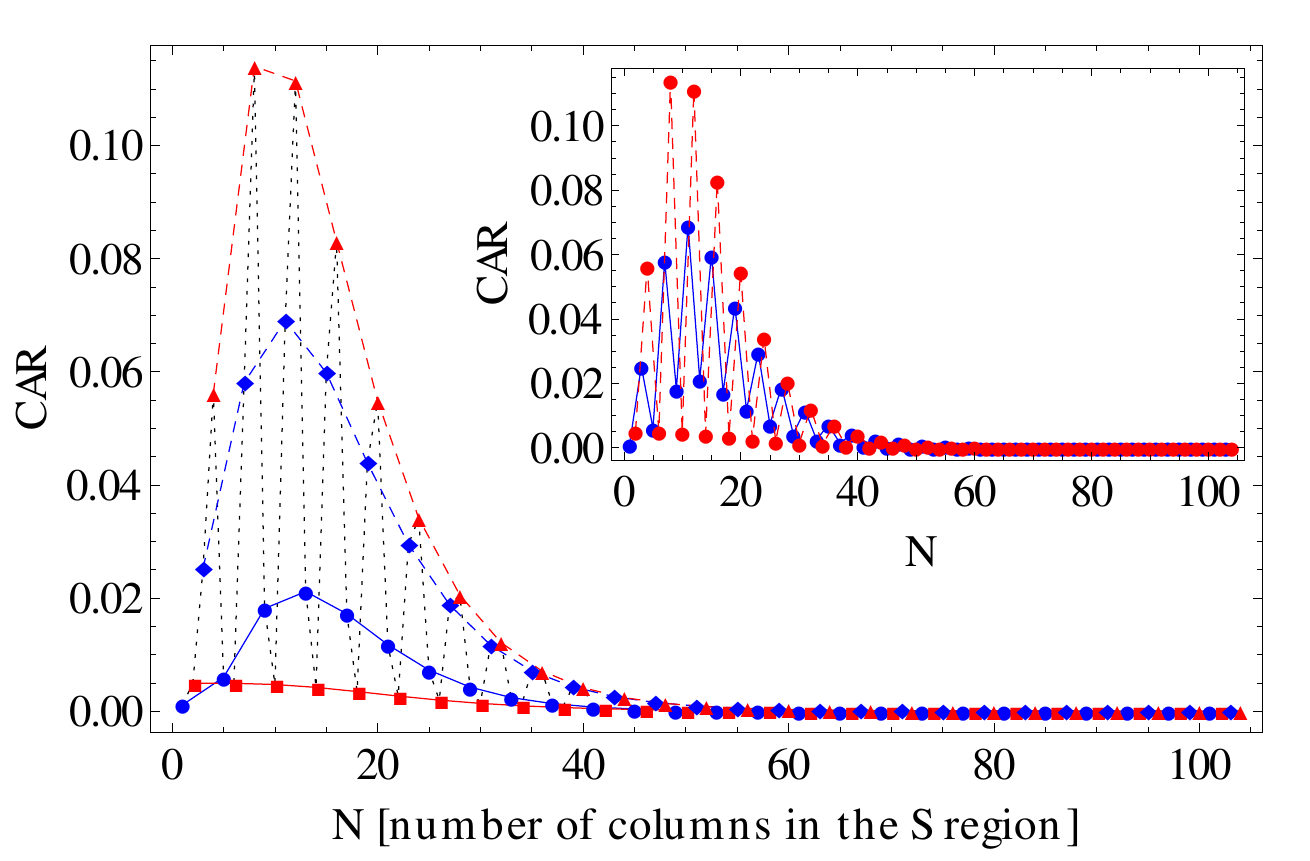}
\caption{(Color online) CAR probability for $E_F = 0$, $V_S = 1.41 t$, $\Delta = 0.1 t$, $\varepsilon = 0.03t$. For this value of $V_S$, $\mathfrak{Re}(k_S a_T) \simeq \pi$. The CAR signal (dotted line) interpolates between four curves: Long-wave length oscillations are suppressed and configurations (1,1) -- circles -- and (1,3) -- rhombi -- as well as (1,2) -- squares -- and (1,4) -- triangles -- cease to be equivalent. The inset shows the same signal, highlighting the two intertwined fast oscillations (see text). }
\label{Fig:extension3}
\end{figure}

We nevertheless highlight a special situation, plotted in Fig.~\ref{Fig:extension3}. The chosen value of $V_S$ implies $\mathfrak{Re}(k_S a_T) \simeq \pi$. This has a very interesting consequence. A look at Eq.~\eqref{eq:transmission_graphene} allows us to infer that, for a given configuration of the interfaces -- keep $i$ and $j$ fixed and vary $L$ --, the signal does not oscillate with $L$, and in that sense, the situation is actually closer to that of Fig.~\ref{Fig:extension0}. The fast oscillations are only between different configurations. However, the equivalence between configurations 1 and 3, as well as between 2 and 4 is lifted. Indeed the CAR signal clearly interpolates between four different curves in Fig.~\ref{Fig:extension3}, contrary to, for instance, the case of Fig.~\ref{Fig:extension2}. The transmission for the configuration (1,3) can be deduced from Eq.~\eqref{eq:transmission_graphene} by taking $i=j=1$ but replacing $L$ by $L+1/2$ (remember that 1 and 3 are related by a translation of half the unit-cell). Now if $\mathfrak{Re}(k_S a_T)$ was strictly $\pi$, the matrix in Eq.~\eqref{eq:transmission_graphene}, for the configuration (1,3) would be $\sigma_z$ and not the identity as for the configuration (1,1). This was exactly what was happening in the 1D case, as one can see in Eqs.~\eqref{eq:alpha_tilde_1} and \eqref{eq:sigma_z}. Of course, the CAR does not vanish exactly because particle-hole symmetry is obviously broken, however the signal shows two intertwined very strong oscillations, as visible in the inset of Fig.~\ref{Fig:extension3}.

\subsubsection{Effects of smooth NS interfaces}

\begin{figure}[h!]
\centering
\includegraphics[width=8cm,clip]{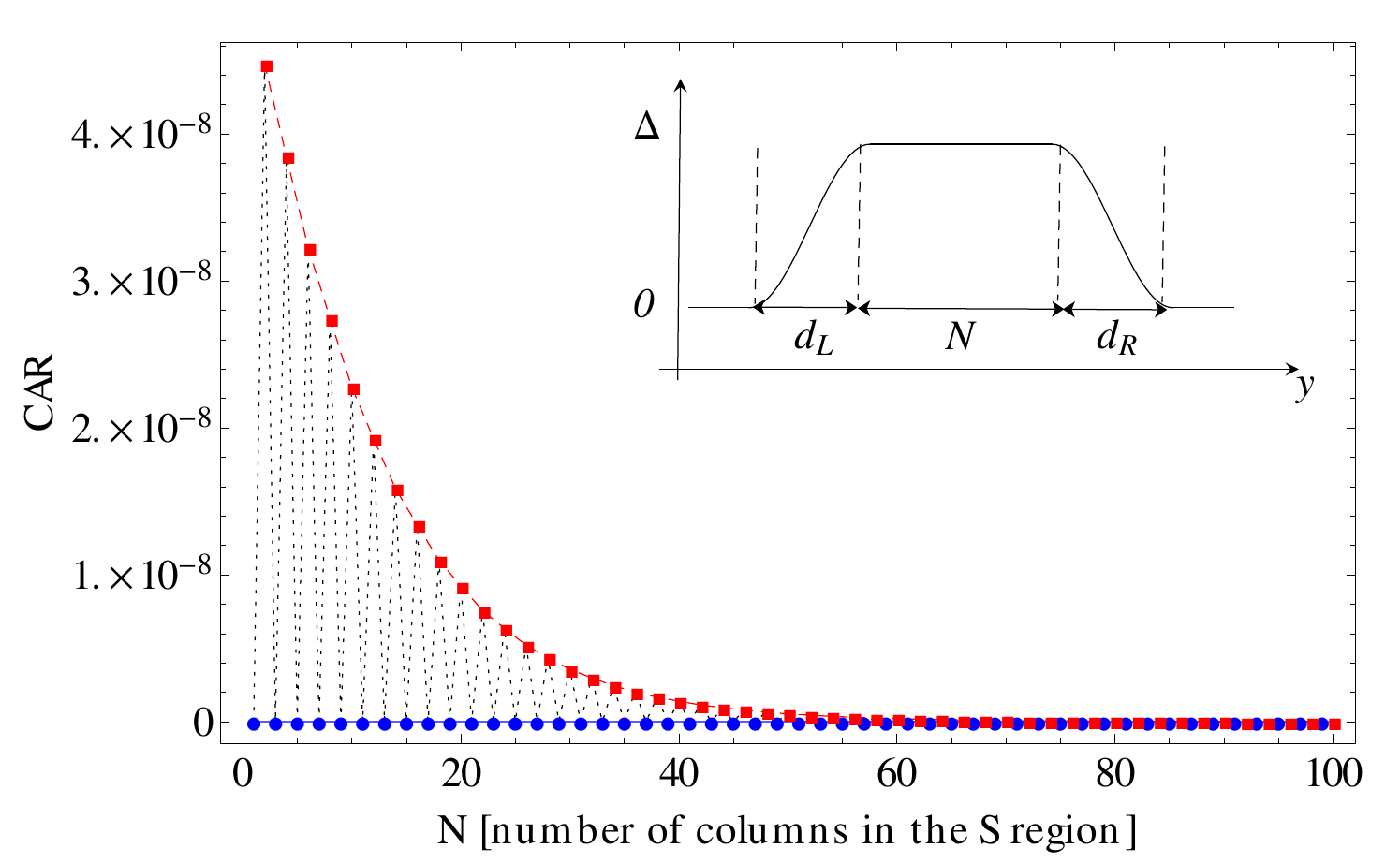}
\caption{(Color online) CAR probability for $E_F = 0$, $V_S = 0 $, $\Delta = 0.1 t$, $\varepsilon = 0.03t$. The superconducting potential is smoothed over a distance $d_R = d_L = 9a_0$ -- with $a_0$ the distance between carbon atoms -- corresponding to three unit-cells. The even-odd effect is preserved.}
\label{Fig:extension4}
\end{figure}

\begin{figure}[h!]
\centering
\includegraphics[width=8cm,clip]{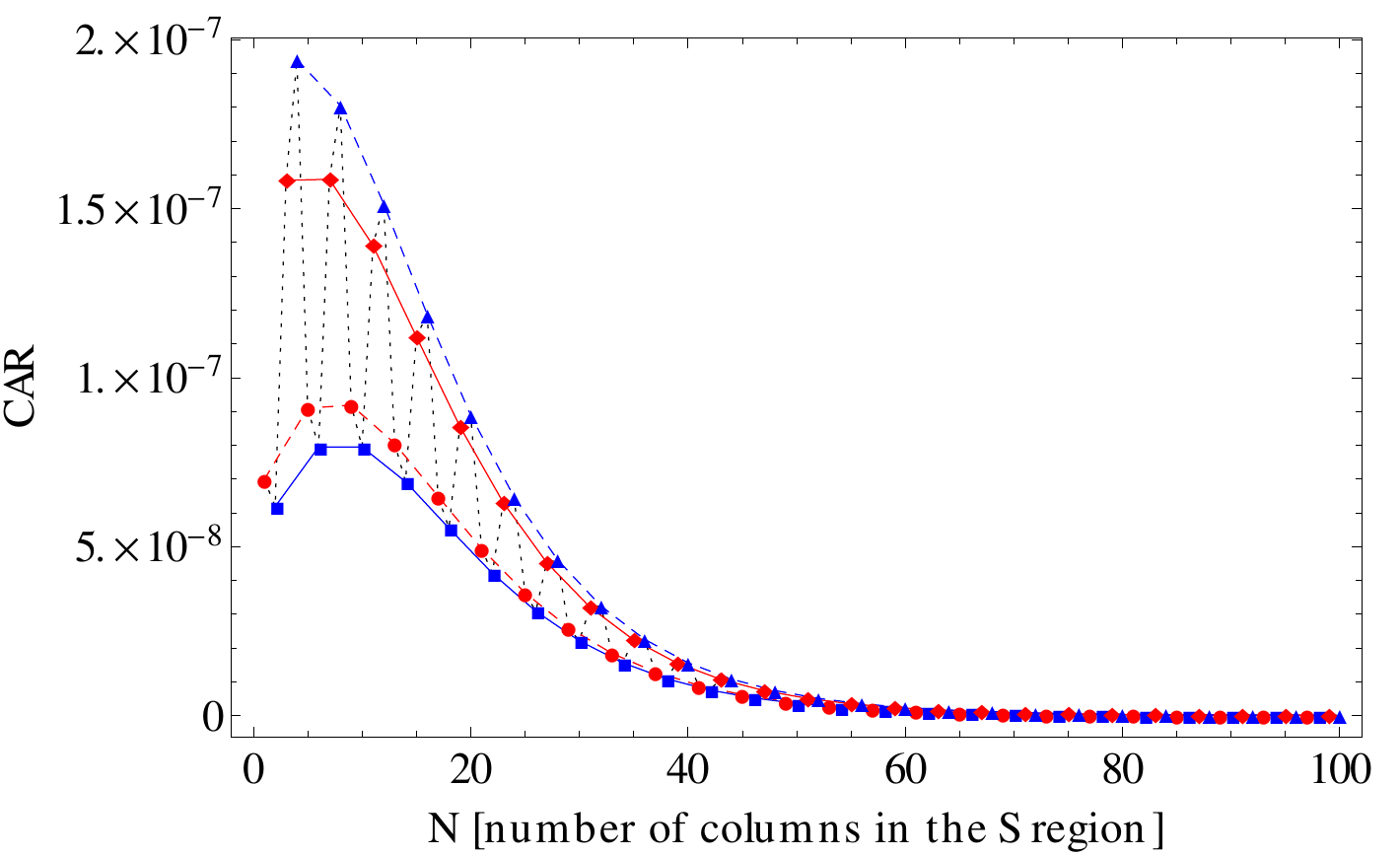}
\caption{(Color online) CAR probability for $E_F = 0$, $V_S = 0 $, $\Delta = 0.1 t$, $\varepsilon = 0.03t$, $d_L = 9a_0$, and $d_R = 18a_0$. The left-right symmetry of the junction is broken. The signal (dotted black line) oscillates between four configurations showed with different symbols (squares, circles, rhombi and triangles). }
\label{Fig:extension4b}
\end{figure}

\begin{figure}[h!]
\includegraphics[width=8cm,clip]{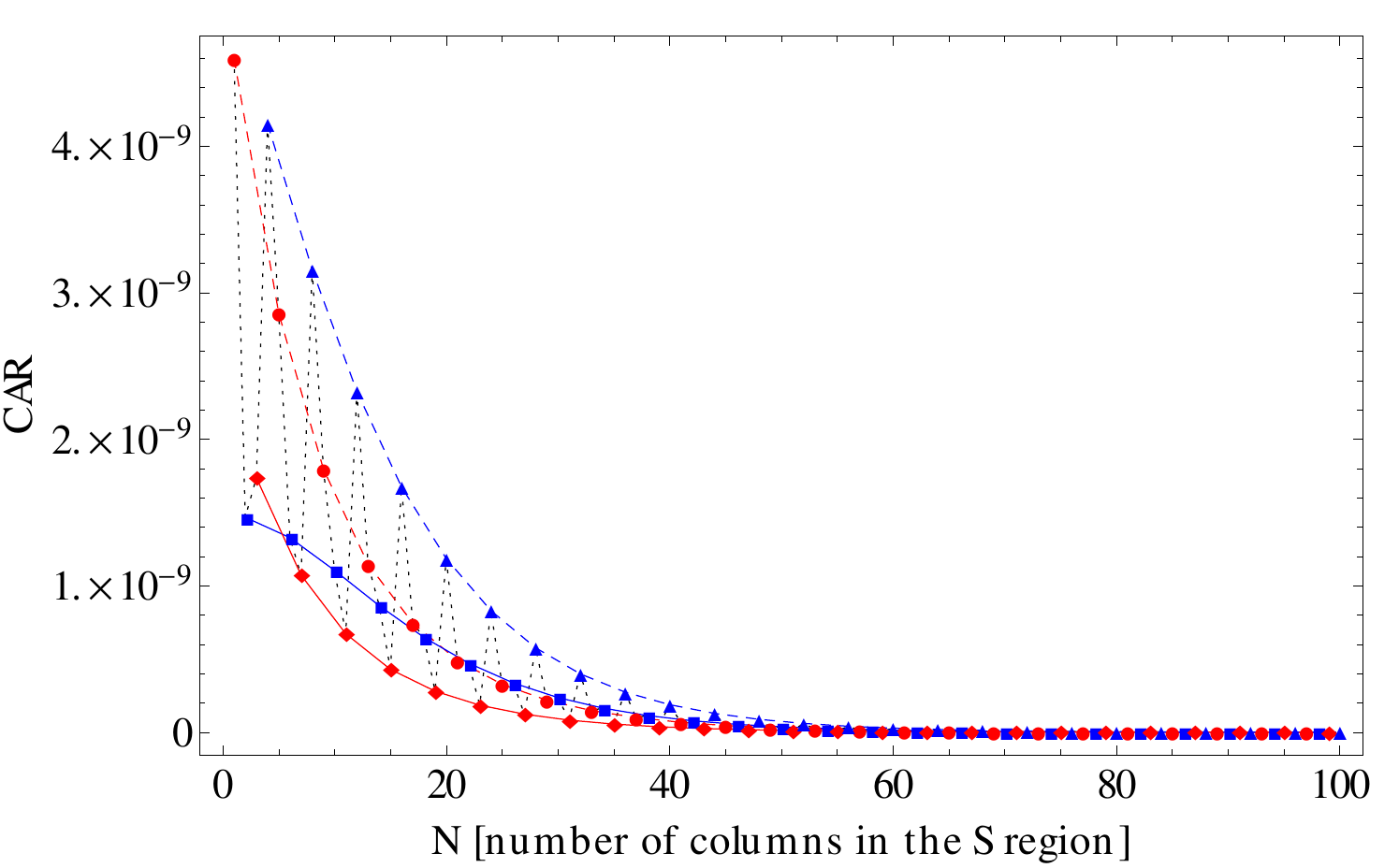}
\caption{(Color online) CAR probability for $E_F = 0$, $V_S = 0 $, $\Delta = 0.1 t$, $\varepsilon = 0.03t$, $d_L = 10a_0$, and $d_R = 10a_0$. The left-right symmetry of the junction is also broken, as the smoothing length is incommensurate with the unit-cell length.}
\label{Fig:extension4c}
\end{figure}

We now investigate the effect of smooth NS interfaces. Each plot in this section corresponds to a situation where the pairing potential goes smoothly from its bulk value $\Delta$ to zero in the normal leads over a length $d_R$ (resp. $d_L$) on the right interface (resp. left interface), as sketched in Fig.~\ref{Fig:extension4}. The smoothing function is defined as $0.5\left(\cos((y-N)/d_R)+1\right)$ on the right interface and $0.5\left(\cos(y/d_L)+1\right)$ on the left one. $d_R$ and $d_L$ are of the order of the coherence length in the superconductor, $\xi = \hbar v_F/\Delta$. The signal is plotted against $N$, the length of the superconducting region over which the gap is constant (see the inset of Fig.~\ref{Fig:extension4}). We have made a numerical calculation of the scattering matrix, using a recursive Green's function algorithm~\cite{Lee81,MacKinnon85,Todorov94,Lake97,Schelter10}. We highlight a few possible effects. At zero Fermi energy the CAR probability is strongly suppressed, as can be seen in Figs.~\ref{Fig:extension4}, \ref{Fig:extension4b} and \ref{Fig:extension4c}. Indeed, as one makes the NS interfaces smoother, the system should be better described by the Dirac-Bogoliubov-de Gennes equations with a pairing potential smooth at the level of the sublattice. In that case, CAR would be completely suppressed at zero $E_F$, due to the orthogonality of the spinor wave-functions of incoming electrons and reflected holes. However, we find that the even-odd effect is preserved in the case of a symmetric smoothing, that is $d_R = d_L$ and $d_R$ is commensurate with the length of the unit-cell, as can be seen in Fig.~\ref{Fig:extension4}. We obtain that the zeros of the CAR probability are lifted by breaking the left-right symmetry of the NSN junction, which can be done either by taking $d_R \neq d_L$, as in Fig.~\ref{Fig:extension4b}, or by choosing a smoothing length that is incommensurate with the unit-cell length, as in Fig.~\ref{Fig:extension4c}.

\noindent These two possibilities share a common feature: As one increases the length of the S region, the signal interpolates between four distinct curves, signaling four inequivalent configurations.\\

\begin{figure}[h!]
\centering
\includegraphics[width=8cm,clip]{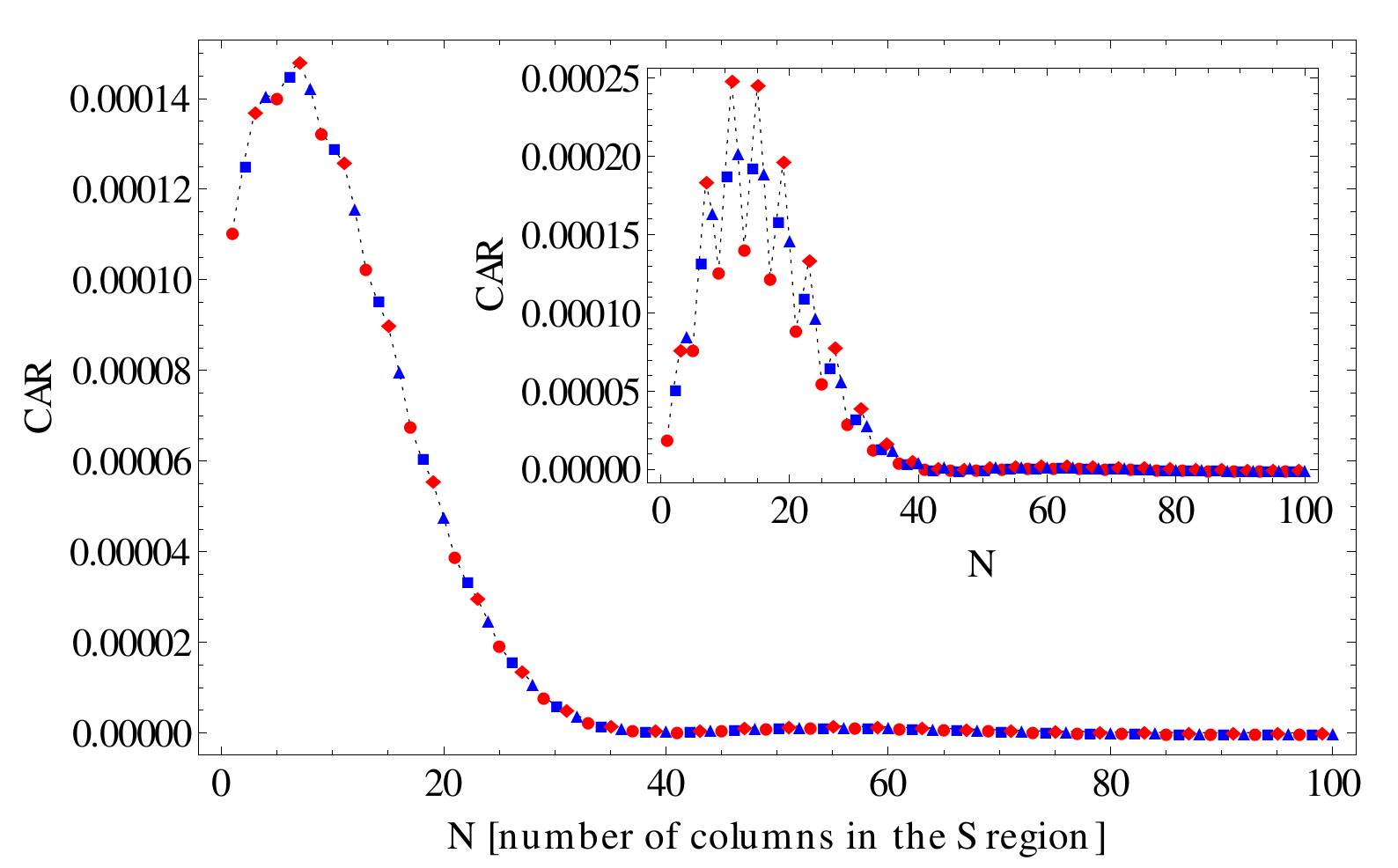}
\caption{(Color online) CAR probability for $E_F = 0$, $V_S = 0.1 t$, $\Delta = 0.1 t$, $\varepsilon = 0.03t$, $d_L = 9a_0$, and $d_R = 9a_0$. The four configurations that were visible in Figs.~\ref{Fig:extension4b} and \ref{Fig:extension4c} now lie practically on top of each other. In the inset we have taken $d_R = d_L = 3a_0$. Fast oscillations between different configurations are more visible.  }
\label{Fig:extension5a}
\end{figure}

\begin{figure}[h!]
\centering
\includegraphics[width=8cm,clip]{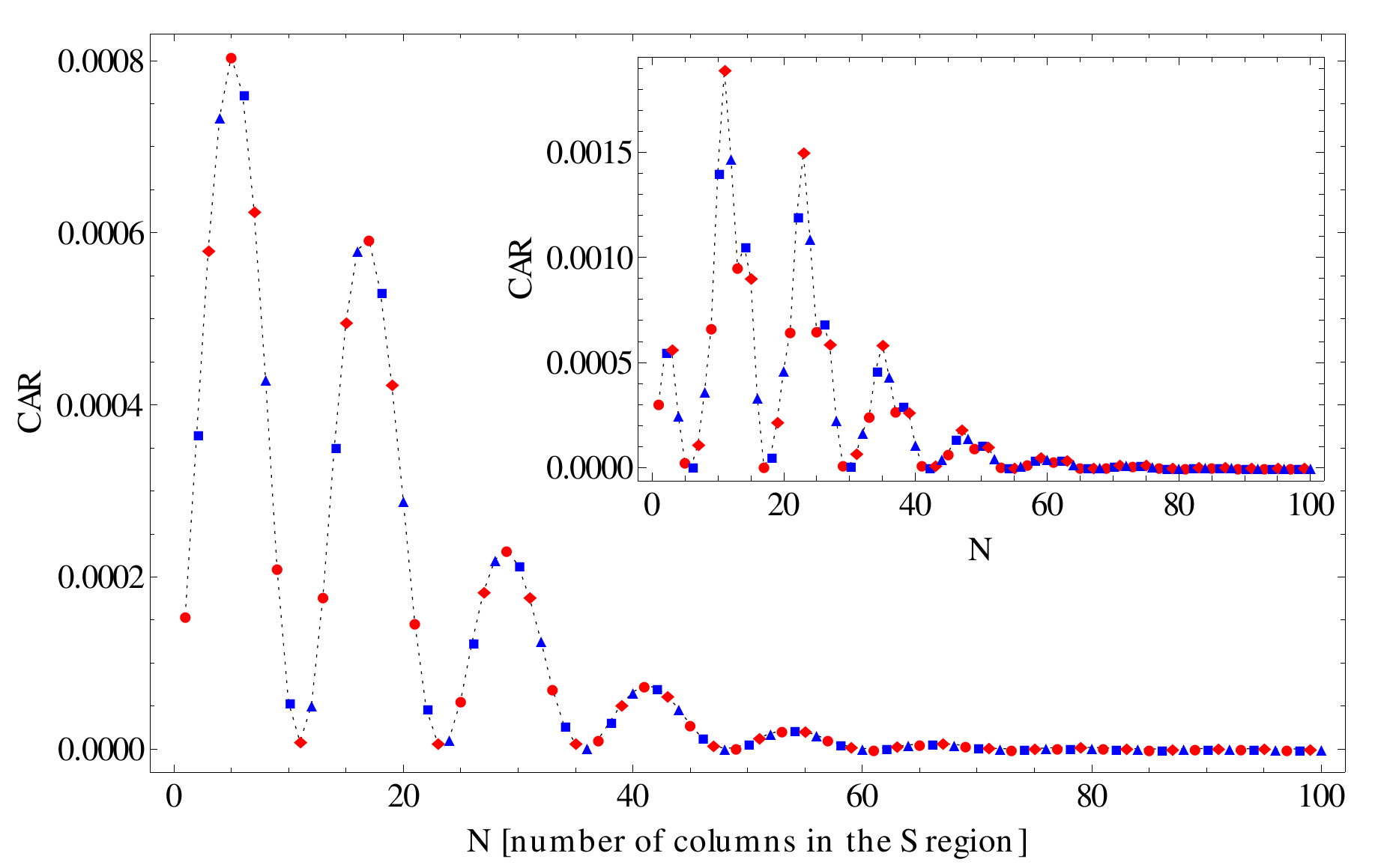}
\caption{(Color online) CAR probability for $E_F = 0$, $V_S = 0.5 t$, $\Delta = 0.1 t$, $\varepsilon = 0.03t$, $d_L = 9a_0$, and $d_R = 9a_0$. We still use different symbols for the four configurations of Figs.~\ref{Fig:extension4b} and \ref{Fig:extension4c}, although the signal interpolates smoothly between them (see text). In the inset we have taken $d_R = d_L = 3a_0$. Fast oscillations between different configurations are more visible. }
\label{Fig:extension5b}
\end{figure}

Next, we combine smoothing and doping by considering the case of finite doping in the S region, while preserving the left-right symmetry. We picked two different values of doping, $V_S = 0.1 t$ and $V_S = 0.5 t$. The corresponding Figs.~\ref{Fig:extension5a} and \ref{Fig:extension5b} are to be contrasted with Fig.~\ref{Fig:extension2}. Note that the Fermi energy goes smoothly from $V_S$ in the S region to 0 in the normal leads on the same distance than the pairing gap. Finite doping seems to smooth out every trace of the even-odd effect as there is hardly any difference between possible configurations of the interfaces. In Fig.~\ref{Fig:extension5a}, the four curves (obtained by increasing the length of the S region by one unit-cell starting from the four a priori different configurations) are still distinct but their interpolation is so smooth that the sublattice hardly matters. This is even clearer for the larger value of doping in Fig.~\ref{Fig:extension5b}, where one cannot distinguish between any sublattice configurations, and the system is better described by the Dirac equation. One can check that by decreasing the smoothing length, the oscillations due to the lattice reappear, as shown in the inset of Figs.~\ref{Fig:extension5a} and \ref{Fig:extension5b}. 

\begin{figure}[h!]
\centering
\includegraphics[width=8cm,clip]{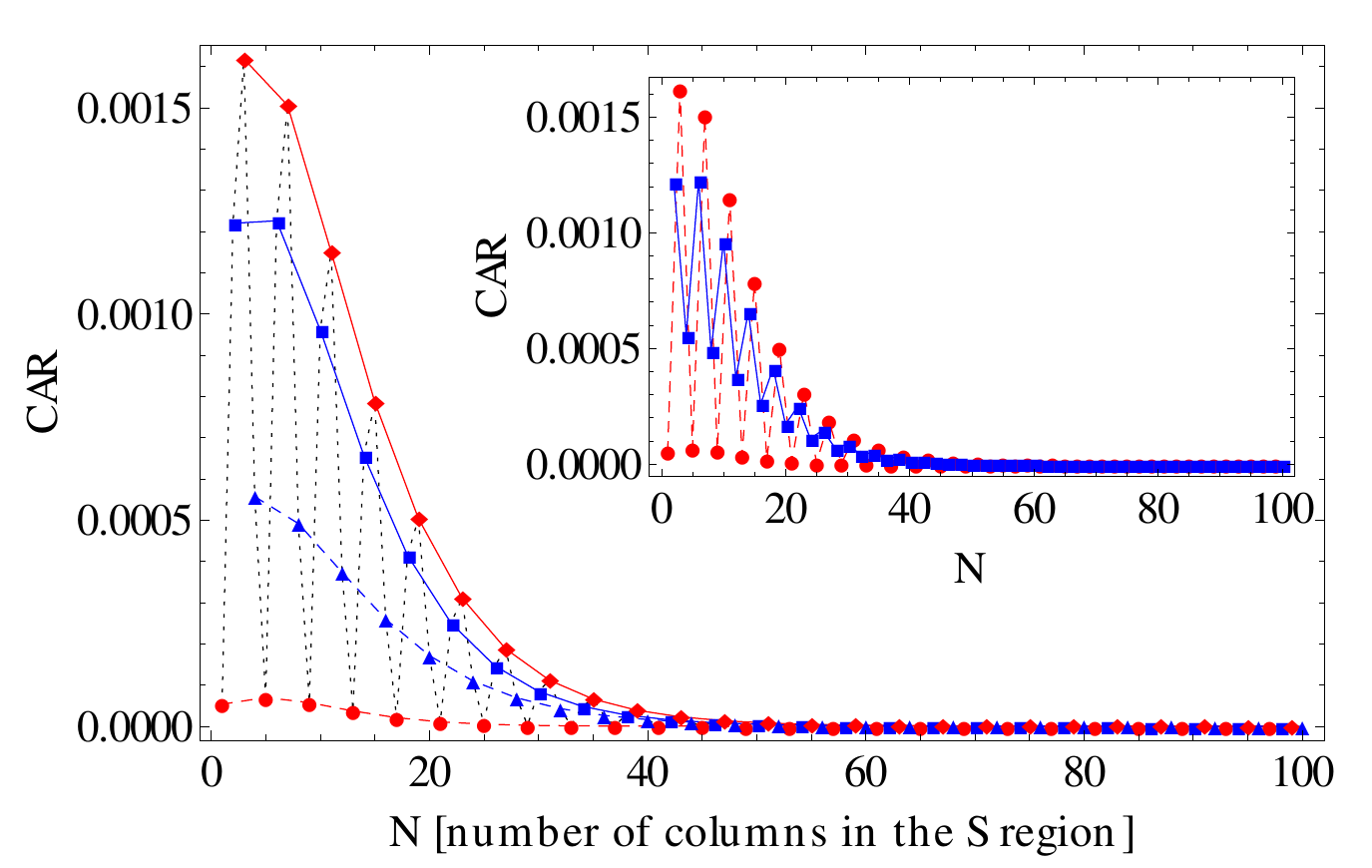}
\caption{(Color online) CAR probability for $E_F = 0$, $V_S = 1.41 t$, $\Delta = 0.1 t$, $\varepsilon = 0.03t$, $d_L = 9a_0$, and $d_R = 9a_0$.  For this value of $V_S$, $\mathfrak{Re}(k_S a_T) \simeq \pi$. The four configurations, showed with different symbols, are again clearly visible. The inset is the same signal, highlighting the two intertwined fast oscillations (see text). }
\label{Fig:extension6}
\end{figure}

However, the special case highlighted in Fig.~\ref{Fig:extension3} is not altered by smoothing -- although the amplitude of the signal is damped -- as is shown in Fig.~\ref{Fig:extension6}. This is a case where oscillations strictly due to doping are suppressed and fast oscillations due to the positions of the interfaces with respect to the sublattice re-emerge. We still identify two intertwined very fast oscillations, {\color{red} as in Fig.~\ref{Fig:extension3} }.

\newpage

\subsubsection{{Irregular NS interfaces}}

\begin{figure}[h!]
\centering
\includegraphics[width=8cm,clip]{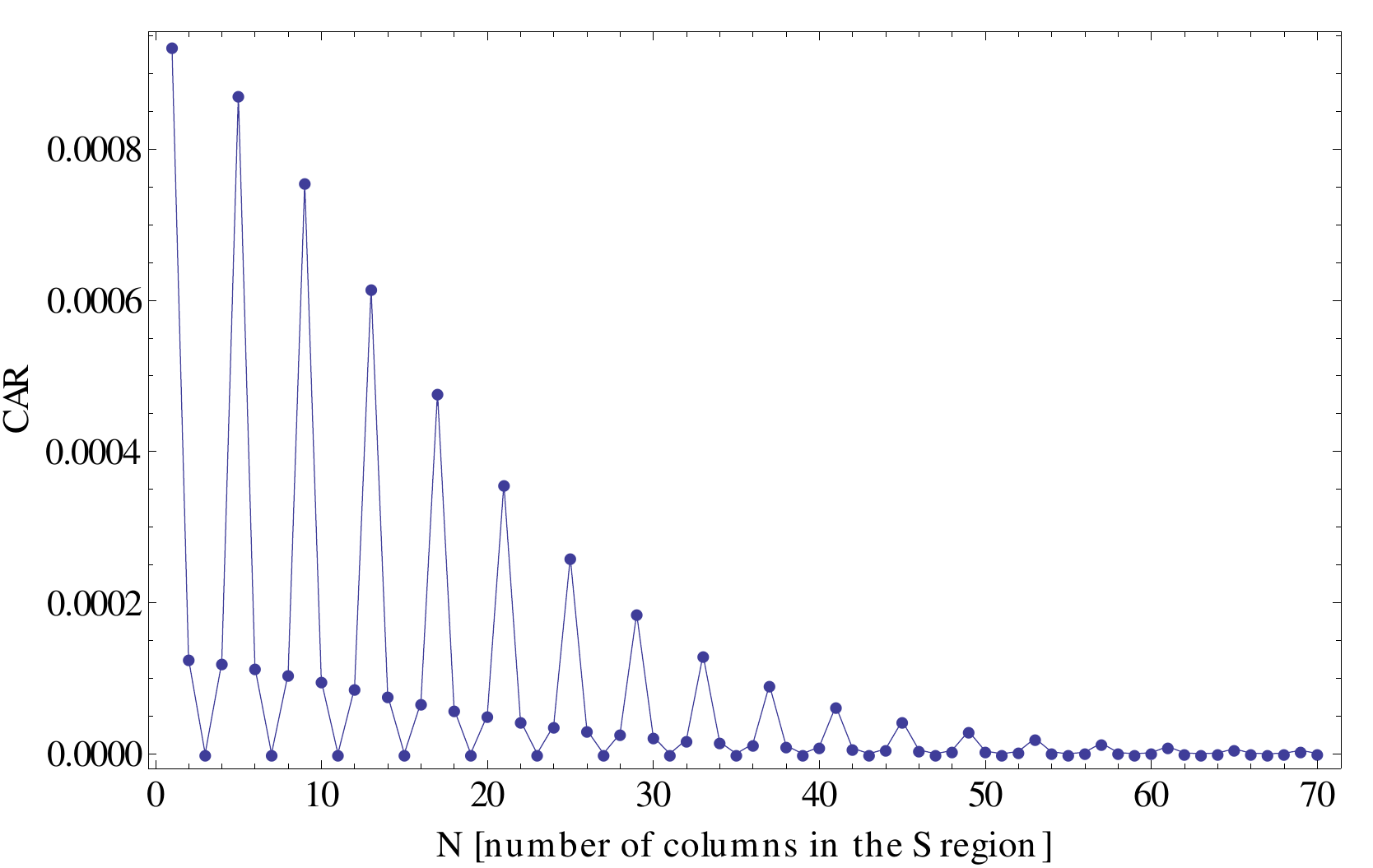}
\caption{(Color online) CAR probability for $E_F = 0 = V_S$, $\Delta = 0.1 t$, $\varepsilon = 0.03t$, for an irregular interface in the transverse direction, with $\omega = 0.01 \pi/a_0$ and $\varphi = 0.513 \pi$ (see text). }
\label{Fig:rough}
\end{figure}

{ In studying smoothed-out interfaces, we addressed the possibility that in real systems, the proximity effect, that is, the tunneling of Cooper pairs between the superconductor and the graphene sheet, exists also away from the superconductor edges. However, we have always assumed the pairing potential not to depend on the position along the transverse direction. In the opposite case, such variations will induce scattering between different transverse modes. As our explanation of the even-odd effect relies on the assumption of independent modes, we expect such a perturbation to lift the exact zeroes. To investigate this situation, we have computed, within the recursive Green's function method, the CAR probability in the following simple model. We have assumed that the right and left sharp NS interfaces are subjected to an envelope function of the form, $f_R(x) = \cos^2(\omega x )$ and $f_L(x) = \cos^2(\omega x + \varphi)$, respectively. $x$ is the coordinate parallel to the interface. As plotted in Fig.~\ref{Fig:rough}, the CAR signal does not show the even-odd effect anymore. However, fast oscillations, on the scale of the unit-cell remain.  }

\subsection{zigzag edges}

\begin{figure}[h!]
\centering
\includegraphics[width=8cm,clip]{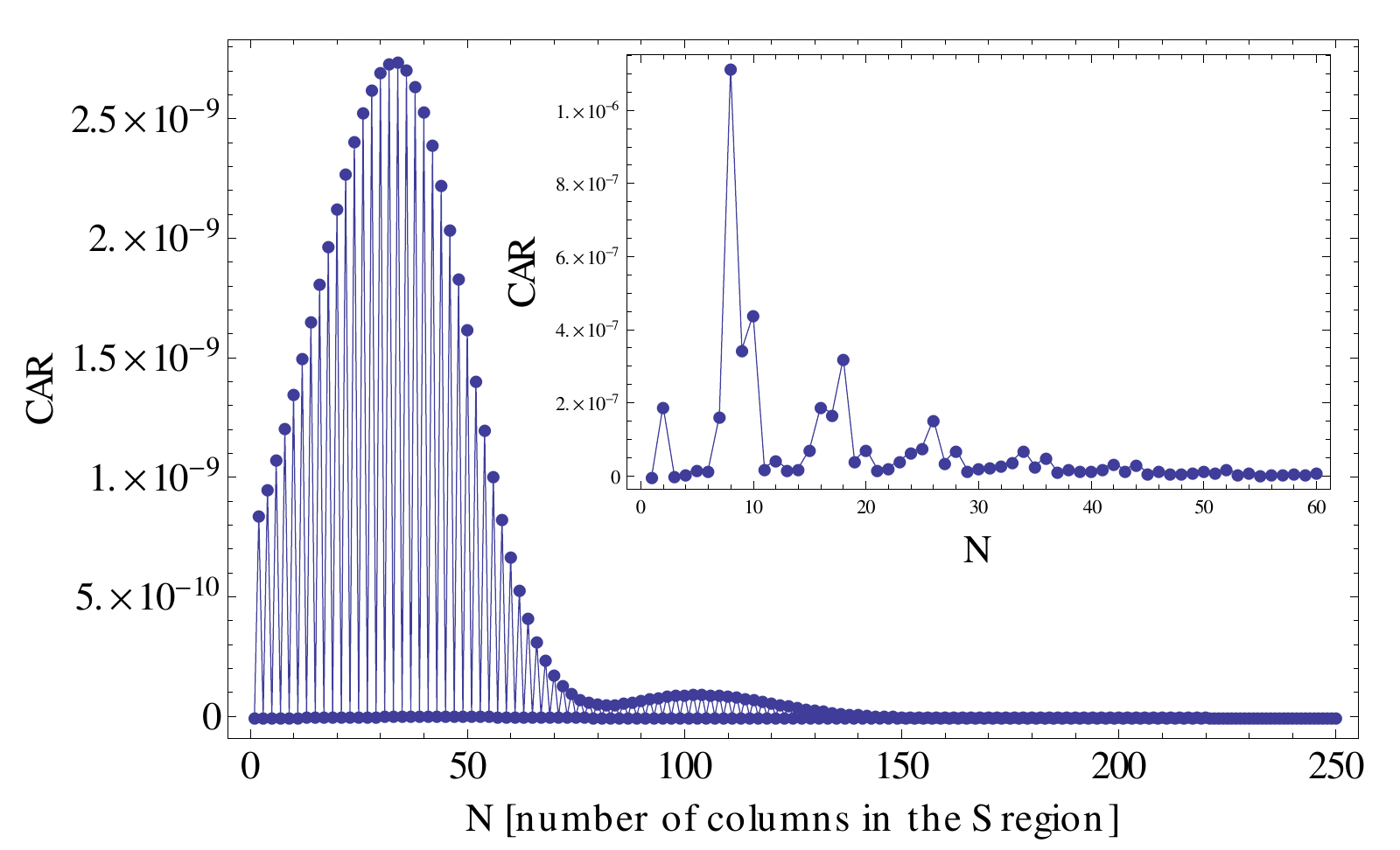}
\caption{(Color online) CAR probability in an anti-zigzag ribbon, for $E_F = 0 = V_S = 0$, $\Delta = 0.0031 t$ and $\varepsilon = 0.003 t$, $W=3$ (number of chains), and $d_R = d_L = 20 \sqrt{3} a_0$ (20 unit-cells). The inset shows the CAR probability for a wider ribbon with $W=21$, all other parameters kept equal.  }
\label{Fig:zz2}
\end{figure}

{ We now briefly turn  to ribbons with zigzag edges and show how the even-odd effect is altered by doping and smoothing. The effect of doping is similar to the armchair case: The CAR probability exhibits long wave-length oscillations, yet the even and odd configurations remain clearly visible, as shown in Fig.~\ref{Fig:zz3}. \\

\begin{figure}[h!]
\centering
\includegraphics[width=8cm,clip]{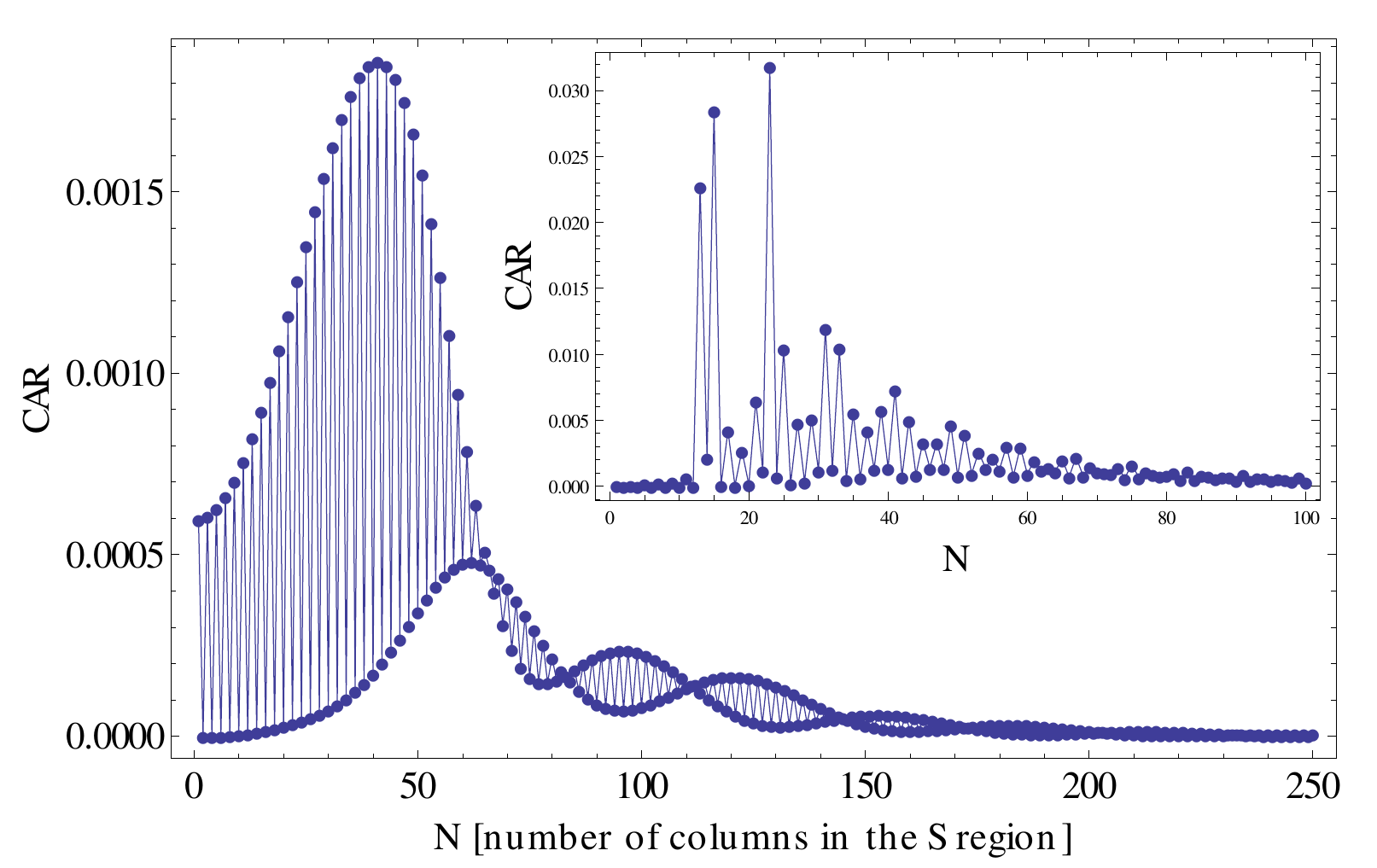}
\caption{(Color online) CAR probability in an anti-zigzag ribbon, for $E_F = 0$, $V_S = 0.001 t$, $\Delta = 0.0031 t$ and $\varepsilon = 0.003 t$, $W=3$ (number of chains). The inset shows the CAR probability for a wider ribbon with $W=21$, all other parameters kept equal.  }
\label{Fig:zz3}
\end{figure}

The effect of smoothing however is quite different and depends strongly on the width of the ribbon, as shown in Fig.~\ref{Fig:zz2}. Similar to the armchair case, we find that the CAR probability is strongly suppressed at zero Fermi energy. The even-odd effect is preserved for narrower ribbons while it is lifted as the width increases. }

\section{Conclusion}

We have investigated the variations of the CAR probability as a function of the length of the superconducting region, and showed that, in a variety of situations, large amplitude oscillations on the scale of the lattice are present. We have discovered conditions under which the CAR probability vanishes altogether only for even lengths of the S region, and explained this effect on the basis of symmetries of the transfer matrix. These sufficient conditions are not strictly tied to a specific lattice and we found that they are satisfied at least in both the simple 1D lattice and graphene nanoribbons. 

Although perturbations tend to lift the perfect zeroes -- for instance, by breaking the left-right symmetry or the particle-hole symmetry -- the CAR amplitude often retains fast oscillations, a remnant of the even-odd effect. It is worth emphasizing that, although our analytical proof of the effect relies on sharp NS interfaces, smoothing of the interfaces, as long as the left-right symmetry is not broken, preserves the effect, as we have shown numerically for graphene. However, the amplitude of the signal is very much reduced, as one approaches the Dirac regime, where selection rules would constrain the CAR probability to be exactly zero in the absence of doping. When doping in the S region is added to the smooth boundaries, in armchair nanoribbons the damping is reduced and fast oscillations tend to disappear, as the smoothing distance approaches and exceeds the coherence length. However, we exhibit a point of very high doping in the S region where the fast oscillations are barely affected by smoothing. Although there is no strict even-odd effect in that case -- the CAR probability does not vanish exactly -- the large amplitude of the oscillations is readily understood on the basis of the transfer matrix. It makes yet another connection with the 1D lattice, and hints at the possibility of finding similar effects in seemingly different lattices.    \\

\section*{ACKNOWLEDGMENTS}
We acknowledge useful discussions with \.Inan\c c Adagideli and Dominik Rüttiger as well as financial support from the DFG priority program "Graphene", the ESF Eurographene and the EU FP7-project SE2ND.

\vspace{1cm}

\appendix

\section{Transfer matrices and wave-matching}
\label{sec:appendix2}

\subsection{General properties of transfer matrices}
The transfer matrix and the scattering matrix of an NSN junction are $4\times4$ matrices of the form
\beq
\mathbf{S} =
\begin{pmatrix}
\mathbf{r} & \mathbf{t}' \\
\mathbf{t} & \mathbf{r}'
\end{pmatrix}, \; \; \; \mathbf{M} =
\begin{pmatrix}
{\bm \alpha} & {\bm \beta} \\
{\bm \gamma} & {\bm \delta}
\end{pmatrix},
\eeq
where $\mathbf{r}, \mathbf{t}, \mathbf{t}', \mathbf{r}'$ are the usual reflection and transmission matrices, from both sides of the barrier. These  $2\times2$ matrices are related to the blocks of $\mathbf{M}$ through~\cite{MelloBook}
\begin{align}
\mathbf{r} &= -{\bm \delta}^{-1}{\bm \gamma}, &\mathbf{r}'&=  {\bm \beta}{\bm \delta}^{-1}, \label{eq:reflection_matrices} \\
\mathbf{t} &= ({\bm \alpha}^\dagger)^{-1}, &\mathbf{t}'&= {\bm \delta}^{-1}.
\end{align}
In the case of time-reversal symmetry, then ${\bm \delta} = {\bm \alpha}^*$ and ${\bm \gamma} = {\bm \beta}^*$. In Eq.~\eqref{eq:mathcalA}, we have introduced the reflection matrices $\mathbf{r}_R$ and $\mathbf{r}_L'$ at the NS interfaces of Fig.~\ref{Fig:NSN_3b}. From Eq.~\eqref{eq:reflection_matrices} we have $\mathbf{r}_R = -{\bm \delta}_R^{-1}{\bm \gamma_R}$ and $\mathbf{r}_L'=  {\bm \beta_L}{\bm \delta}_L^{-1}$. Using inversion formulas and $\mathbf{M}_L(0) = [\mathbf{M}_R(0)]^{-1}$ it follows that $\mathbf{r}_L'=  - {\bm \alpha}_R^{-1}{\bm \beta}_R$. This proof also holds for armchair graphene nanoribbons.\\

\subsection{Important relations in the case of the 1D lattice}

We write down explicitly the wave-matching equations for the 1D lattice and prove important relations between transfer matrices.\\

Let us first prove Eq. \eqref{eq:M_5}, $\mathbf{M}_L(0) = [\mathbf{M}_R(0)]^{-1}$. The transfer matrix $\mathbf{M}_R(0)$ is defined through the wave-matching equations in Eq. \eqref{eq:wave-matching_1}, which explicitly read, for $\varepsilon > 0$,
\begin{widetext}
\begin{align}
&a_{1,e} + b_{1,e} &= &b_{S,1}F_+(k_S) + b_{S,2}F_+(k_S)^* + a_{S,1}F_+(k_S) + a_{S,2}F_+(k_S)^*\;, \nn \\
&a_{1,h} + b_{1,h} &= &b_{S,1} + b_{S,2} + a_{S,1} + a_{S,2}\;, \nn \\
&a_{1,e}e^{ik_e a} + b_{1,e}e^{-ik_e a} &= &b_{S,1}F_+(k_S)e^{ik_S a} + b_{S,2}F_+(k_S)^*e^{-ik_S^* a} + a_{S,1}F_+(k_S)e^{-ik_S a} + a_{S,2}F_+(k_S)^*e^{ik_S^*a}\;, \nn \\
&a_{1,h}e^{-ik_h a} + b_{1,h}e^{ik_h a} &= &b_{S,1}e^{ik_S a} + b_{S,2}e^{-ik_S^* a} + a_{S,1}e^{-ik_S a} + a_{S,2}e^{ik_S^* a}\;,
\end{align}
\end{widetext}
where $F_+$ was defined in Eq.~\eqref{eq:F}.

Similarly, $\mathbf{M}_L(0)$ is defined through the equations 
\begin{widetext}
\begin{align}
&b_{2,e} + a_{2,e} &= &b_{S,1}F_+(k_S) + b_{S,2}F_+(k_S)^* + a_{S,1}F_+(k_S) + a_{S,2}F_+(k_S)^* \nn \\
&b_{2,h} + a_{2,h} &= &b_{S,1} + b_{S,2} + a_{S,1} + a_{S,2} \nn \\
&b_{2,e}e^{ik_e a} + a_{2,e}e^{-ik_e a} &= &b_{S,1}F_+(k_S)e^{ik_S a} + b_{S,2}F_+(k_S)^*e^{-ik_S^* a} + a_{S,1}F_+(k_S)e^{-ik_S a} + a_{S,2}F_+(k_S)^*e^{ik_S^*a} \nn \\
&b_{2,h}e^{-ik_h a} + a_{2,h}e^{ik_h a} &= &b_{S,1}e^{ik_S a} + b_{S,2}e^{-ik_S^* a} + a_{S,1}e^{-ik_S a} + a_{S,2}e^{ik_S^* a} \label{eq:app_M_L}
\end{align}
\end{widetext}
as is clear from Fig.~\ref{Fig:NSN_2}. Inverting these two systems we would get 
\beq
\begin{pmatrix} \mathbf{B}_S \\ \mathbf{A}_S \end{pmatrix} = \mathbf{M}_R(0) \begin{pmatrix}  \mathbf{A}_1 \\ \mathbf{B}_1  \end{pmatrix}, 
\eeq
and
\beq
\begin{pmatrix} \mathbf{B}_2 \\ \mathbf{A}_2 \end{pmatrix} = \mathbf{M}_L(0) \begin{pmatrix}  \mathbf{B}_S \\ \mathbf{A}_S  \end{pmatrix}. 
\eeq
respectively. However one can check by inspection that, for given $\mathbf{B}_S$ and $\mathbf{A}_S$, $\mathbf{B}_2 = \mathbf{A}_1$ and $\mathbf{A}_2 = \mathbf{B}_1$ are still solutions of the system \eqref{eq:app_M_L}. Therefore we have,
\beq
\begin{pmatrix} \mathbf{A}_1 \\ \mathbf{B}_1 \end{pmatrix} = \mathbf{M}_L(0) \begin{pmatrix}  \mathbf{B}_S \\ \mathbf{A}_S  \end{pmatrix}, 
\eeq
and, as a consequence, $\mathbf{M}_L(0) = [\mathbf{M}_R(0)]^{-1}$.\\

We now prove Eq.~\eqref{eq:mathcalA_2}. To do so one should write the system of wave-matching equations for the transfer matrix $\hat{\mathbf{M}}_L(0)$ defined in Fig.~\ref{Fig:NSN_3b}. Comparing it to the system defining $\mathbf{M}_R(0)$, one finds the simple relation
\beq
\hat{\mathbf{M}}_L(0) = \begin{pmatrix}
0 & \mathbf{1}_2 \\
\mathbf{1}_2 & 0
\end{pmatrix} \left[\mathbf{M}_R(0)\right]^{-1} \begin{pmatrix}
0 & \mathbf{1}_2 \\
\mathbf{1}_2 & 0
\end{pmatrix}, \label{eq:M_6b}
\eeq
from which follows $\hat{\mathbf{r}}_L' = \mathbf{r}_R$ and then Eq.~\eqref{eq:mathcalA_2}.

\subsection{Important relations in the case of armchair graphene nanoribbons}
The proof of  $\mathbf{M}_L^{(i)}(0) = [\mathbf{M}_R^{(i)}(0)]^{-1}$ is straightforward and utterly similar to the proof for the linear chain. In the present section, we focus on the proof of Eq.~\eqref{eq:r_graphene}, which comes from a relation very similar to Eq.~\eqref{eq:M_6b}. Again we prove the relation by simple inspection. Let us write the wave-matching equations for both $\mathbf{M}_R^{(i)}(0)$ and $\hat{\mathbf{M}}_L^{(i)}(0)$, as defined in Fig.~\ref{Fig:NSN_5b}. To that end we use the scattering states defined in Appendix~\ref{sec:appendix1} for graphene. First for $\mathbf{M}_R^{(i)}(0)$, in the case $E_F = V_S = 0$,
\begin{widetext}
\begin{align}
&a_{1,e}(-se^{-is\theta_e}) + b_{1,e}(-se^{is\theta_e}) &= &b_{S,1} (-e^{-i\theta_S})F + b_{S,2} e^{i\theta_S}F^* + a_{S,1}(-e^{i\theta_S})F + a_{S,2}e^{-i\theta_S}F^* \nn \\
&a_{1,e}e^{-isk_e a_t/2} + b_{1,e}e^{isk_e a_T/2} &= &b_{S,1}Fe^{-ik_S a_T/2} + b_{S,2}F^*e^{-ik_S a_T/2} + a_{S,1}Fe^{ik_S a_T/2} + a_{S,2}F^*e^{ik_S a_T/2} \nn \\
&a_{1,h}(-s'e^{is'\theta_h}) + b_{1,h}(-s'e^{-is'\theta_h}) &= &b_{S,1} (-e^{-i\theta_S}) + b_{S,2} e^{i\theta_S} + a_{S,1}(-e^{i\theta_S}) + a_{S,2}e^{-i\theta_S} \nn \\
&a_{1,h}e^{is'k_h a_t/2} + b_{1,e}e^{-is'k_h a_T/2} &= &b_{S,1}e^{-ik_S a_T/2} + b_{S,2}e^{-ik_S a_T/2} + a_{S,1}e^{ik_S a_T/2} + a_{S,2}e^{ik_S a_T/2} \label{eq:app_graphene_1}
\end{align} 
\end{widetext}
where we defined $F=F_{1+}$ to simplify notations. For $\hat{\mathbf{M}}_L^{(i)}(0)$, still in the case $E_F = V_S = 0$, we have
\begin{widetext}
\begin{align}
&b_{2,e}(-se^{-is\theta_e})e^{isk_e a_t/2} + a_{2,e}(-se^{is\theta_e})e^{-isk_e a_t/2} &= &a_{S,1} (-e^{-i\theta_S})Fe^{ik_S a_T/2} + a_{S,2} e^{i\theta_S}F^*e^{ik_S a_T/2} \nn  \\
& & &+ b_{S,1}(-e^{i\theta_S})Fe^{-ik_S a_T/2}  + b_{S,2}e^{-i\theta_S}F^*e^{-ik_S a_T/2} \nn \\
&b_{2,e} + a_{2,e} &= &a_{S,1}F + a_{S,2}F^* + b_{S,1}F + b_{S,2}F^* \nn \\
&b_{2,h}(-s'e^{is'\theta_h})e^{-is'k_h a_t/2} + a_{2,h}(-s'e^{-is'\theta_h})e^{+is'k_h a_T/2} &= &a_{S,1} (-e^{-i\theta_S})e^{ik_S a_T/2} + a_{S,2} e^{i\theta_S}e^{ik_S a_T/2} + b_{S,1}(-e^{i\theta_S})e^{-ik_S a_T/2} \nn  \\
& & &+ b_{S,2}e^{-i\theta_S}e^{-ik_S a_T/2} \nn \\
&b_{2,h} + a_{2,h} &= &a_{S,1}+ a_{S,2} + b_{S,1}+ b_{S,2} \label{eq:app_graphene_2}
\end{align}
\end{widetext}
It is just a matter of algebra to verify that the transformation,
\begin{widetext}
\beq
\begin{pmatrix} \mathbf{A}_1 \\ \mathbf{B}_1 \end{pmatrix} \rightarrow \begin{pmatrix}
0 & \mathbf{1}_2 \\
\mathbf{1}_2 & 0
\end{pmatrix} \textrm{Diag} \left[
-s e^{-is\theta_e}, -s'e^{is'\theta_h}, -se^{is\theta_e}, -s'e^{-is'\theta_h}
\right] 
\begin{pmatrix} \mathbf{B}_2 \\ \mathbf{A}_2 \end{pmatrix}
\eeq
and
\beq
\begin{pmatrix} \mathbf{B}_S \\ \mathbf{A}_S \end{pmatrix} \rightarrow \begin{pmatrix}
0 & \mathbf{1}_2 \\
\mathbf{1}_2 & 0
\end{pmatrix}  \textrm{Diag} \left[
-e^{i\theta_S}, e^{-i\theta_S}, -e^{-i\theta_S}, e^{i\theta_S}
\right]
\begin{pmatrix} \mathbf{B}_S \\ \mathbf{A}_S \end{pmatrix}
\eeq
\end{widetext}
maps the system \eqref{eq:app_graphene_1} to the system \eqref{eq:app_graphene_2}, from which we deduce 
\begin{align}
\hat{\mathbf{M}}_L^{(1)}(0) = \mathcal{P}_{eh} \begin{pmatrix}
0 & \mathbf{1}_2 \\
\mathbf{1}_2 & 0
\end{pmatrix}
\left[\mathbf{M}_R^{(1)}(0)\right]^{-1} \begin{pmatrix}
0 & \mathbf{1}_2 \\
\mathbf{1}_2 & 0
\end{pmatrix}\mathcal{P}_{S}, \label{eq:M_9}
\end{align}
where we have defined 
\beq
\mathcal{P}_{eh} = \textrm{Diag} \left[
-s e^{-is\theta_e}, -s'e^{is'\theta_h}, -se^{is\theta_e}, -s'e^{-is'\theta_h}
\right]
\eeq and 
\beq \mathcal{P}_{S} = \textrm{Diag} \left[
-e^{i\theta_S}, e^{-i\theta_S}, -e^{-i\theta_S}, e^{i\theta_S}
\right],
\eeq
with $\theta_S = \theta(k_S)$. Equation~\eqref{eq:r_graphene} follows directly.

\section{Scattering states in graphene}
\label{sec:appendix1}

In this appendix, we detail the solutions of the Bogoliubov-de Gennes equations in both normal and superconducting regions. In the normal regions, right ($+$) and left ($-$) going waves are defined for electron-like excitations as
\beq
\Psi^\pm_{e,\ell,m}(k) =  \left(\varphi_{eA}^{\pm}e^{\pm i s k_e y_{\ell,mA}} + \varphi_{eB}^{\pm}e^{\pm i s k_e y_{\ell,mB}} \right),
\eeq
with
\beq
\begin{pmatrix}
\varphi_{eA}^{\pm}\\
\varphi_{eB}^{\pm}
\end{pmatrix} =  \begin{pmatrix}
-s e^{\mp is\theta_e} \\
1
\end{pmatrix},
\eeq
$s = \textrm{sgn}(\varepsilon + E_F)$ and $\theta_e = \theta(k_e)$. We have dropped the sine function, since it simplifies in the wave-matching conditions. For hole-like excitations
\beq
\Psi^\pm_{h,\ell,m}(k) =  \left(\varphi_{hA}^{\pm}e^{\mp i s' k_h y_{\ell,mA}} + \varphi_{hB}^{\pm}e^{\mp i s' k_h y_{\ell,mB}} \right),
\eeq
with
\beq
\begin{pmatrix}
\varphi_{hA}^{\pm}\\
\varphi_{hB}^{\pm}
\end{pmatrix} =  \begin{pmatrix}
-s' e^{\pm is'\theta_h} \\
1
\end{pmatrix},
\eeq
$s' = \textrm{sgn}(-\varepsilon + E_F)$, and $\theta_h = \theta(k_h)$. The wave-vectors $k_e$ and $k_h$ are the positive roots of $\varepsilon = -E_F + E_s(k,p)$ and $\varepsilon = E_F - E_{s'}(k,p)$, respectively. Particle-hole symmetry of the spectrum, $E_F = 0$, implies 
\beq
k_e = k_h. 
\eeq

In the superconducting region, where $\Delta \neq 0$, the BdG Hamiltonian has four eigenvalues labeled $\epsilon_{1\pm} = \pm \sqrt{\Lambda_1(k)^2 + \Delta^2}$ and $\epsilon_{2\pm} = \pm \sqrt{\Lambda_2(k)^2 + \Delta^2}$ with $\Lambda_1(k) = E_+(k,p) - E_F-V_S$ and $\Lambda_2(k) = E_+(k,p) + E_F+V_S$. The corresponding eigenvectors are $\chi_{1\pm}(k) = \left( -e^{-i\theta(k)} F_{1\pm}(k),F_{1\pm}(k),-e^{-i\theta(k)},1  \right)^T$ and $\chi_{2\pm}(k) = \left( e^{-i\theta(k)} F_{2\pm}(k),F_{2\pm}(k),e^{-i\theta(k)},1  \right)^T$ with $F_{1\pm}(k)=
(\Lambda_1(k)\pm  \sqrt{\Lambda_1(k)^2 + \Delta^2})/\Delta$ and $F_{2\pm}(k)=
(-\Lambda_2(k)\pm  \sqrt{\Lambda_2(k)^2 + \Delta^2})/\Delta$. As for the linear chain, there are four available wave-vectors, at a given excitation energy. However, we must distinguish between the cases, $E_F +V_S \neq 0$ and $E_F +V_S = 0$. Indeed, in the former case, all modes stem from one eigenvalue, e.g. $\epsilon_{1+}$ for $\varepsilon>0$ and $E_F+V_S>0$. For $\varepsilon<\Delta$, the wave-vectors are given by
\beq
k = \pm \frac{2}{a_T} \textrm{Arccos}\left(\frac{(E_F +V_S \pm i\sqrt{\Delta^2 -\varepsilon^2})^2-(1+\epsilon_p^2)}{2 \epsilon_p}\right). \label{eq:k_s2}
\eeq
We write these wave-vectors as $k_S$, $-k_S^*$ (right-decaying modes) and $-k_S$, $k_S^*$ (left-decaying modes) where $k_S$ is the solution with positive real and imaginary parts. A general solution of the BdG equations in the S region, for $0<\varepsilon<\Delta$ and $E_F + V_S >0$ is  a superposition of right and left decaying modes of the form
\begin{align}
\Psi_{S,\ell,m} &= b_{S,1} \mathfrak{D}(k_S)\chi_{1+}(k_S)  +  b_{S,2} \mathfrak{D}(-k_S^*) \chi_{1+}(-k_S^ *)  \nn \\
&+ a_{S,1} \mathfrak{D}(-k_S)\chi_{1+}(-k_S) +  a_{S,2} \mathfrak{D}(k_S^*)\chi_{1+}(k_S^ *),
\end{align}
where we have introduced the diagonal matrix $\mathfrak{D}(k) = \textrm{Diag}[e^{iky_{\ell,mA}},e^{iky_{\ell,mB}},e^{iky_{\ell,mA}},e^{iky_{\ell,mB}}]$.  In the case $E_F + V_S = 0$, $\epsilon_{1+}$ and $\epsilon_{2+}$ (resp. $\epsilon_{1-}$ and $\epsilon_{2-}$) are degenerate. There are still four modes but only two different wave-vectors: $k_S$ and $-k_S$, with $\mathfrak{Im}(k_S)>0$. Then, a general solution in the S region, for $0<\varepsilon<\Delta$ and $E_F + V_S = 0$ is of the form
\begin{align}
\Psi_{S,\ell,m} &= b_{S,1} \mathfrak{D}(k_S) \chi_{1+}(k_S)+  b_{S,2} \mathfrak{D}(k_S) \chi_{2+}(k_S) \nn \\
&+ a_{S,1} \mathfrak{D}(-k_S) \chi_{1+}(-k_S) +  a_{S,2} \mathfrak{D}(-k_S) \chi_{2+}(-k_S). 
\end{align}
We emphasize the important constraint from particle-hole symmetry of the spectrum that is 
\beq
\mathfrak{Re}(k_S) = \left\lbrace \begin{array}{ll} 0 &\mbox{if $\epsilon_p<0$}, \\  2\pi/a_T &\mbox{if $\epsilon_p>0$}.   \end{array} \right.
\eeq

\bibliography{Top_ins_wurzburg}

\end{document}